\documentclass[aps,prstper,twocolumn,groupedaddress]{revtex4-2}

\usepackage{graphicx}
\usepackage{xcolor}
\usepackage[normalem]{ulem}

\begin{document}


\title{Prevalence of a growth mindset among introductory astronomy students}



\author{Moire K. M. Prescott}
\email[]{mkpresco@nmsu.edu}
\affiliation{Department of Astronomy, New Mexico State University, Las Cruces, NM 88003, USA}

\author{Laura Madson}
\affiliation{Department of Psychology, New Mexico State University, Las Cruces, NM 88003, USA}

\author{Sandra M. Way}
\affiliation{Department of Sociology, New Mexico State University, Las Cruces, NM 88003, USA}

\author{Kelly N. Sanderson}
\affiliation{Department of Astronomy, New Mexico State University, Las Cruces, NM 88003, USA }


\date{\today}

\begin{abstract}
While many previous studies have indicated that encouraging a growth mindset can improve student learning outcomes, this 
conclusion's applicability to college-level astronomy classrooms 
remains poorly understood owing to the variation in students' 
overall and domain-specific learning attitudes.  To address 
this, we surveyed undergraduate students in an introductory astronomy class about their attitudes towards learning astronomy over the course of five semesters. 
Overall, students felt an affinity for astronomy, felt moderately competent, perceived astronomy to be intermediate in terms of difficulty, and agreed strongly with standard statements reflecting a ``growth mindset,'' i.e., the belief that intelligence is malleable rather than fixed from birth. Their responses were stable over the course of the semester and did not appear to depend strongly on student demographics.  The unexpected start of the COVID-19 pandemic and the associated shift to all-virtual learning correlated with a drop in their affinity for astronomy, a small decrease in their perceived competence, and an increase in the perceived difficulty of the topic. Their overall learning mindset showed negligible change during this time, emphasizing the stability of their belief in a growth mindset as compared to other measured learning attitudes. However, more nuanced questions about their behaviors and interpretations in the classroom, about how they felt ``in the moment'', and about what factors were most important for their success in the class revealed significantly lower alignment with a growth mindset. This suggests that while introductory astronomy students may believe that they have a growth mindset, 
this mindset is not necessarily reflected in
their self-reported classroom behaviors or measured responses to actual learning challenges.

\end{abstract}


\maketitle


\section{Introduction\label{sec:intro}}

Many studies have demonstrated that the belief that one’s intelligence can be changed with time and effort, dubbed a “growth mindset”, leads to improved performance due to the suite of pro-learning behaviors it encourages, e.g., demonstrating higher academic engagement, taking on additional challenges, making plans to improve after a setback, setting learning-oriented goals, and adopting mastery-oriented strategies \citep[e.g.,][]{Dweck1988,Dweck1999,Hong1999,Heine2001,Blackwell2007,Burnette2013,Paunesku2015,Smiley2016} (but see also \citeauthor{Bahnik2017} \citeyear{Bahnik2017}). Conversely, a belief that one’s intelligence is fixed and unchangeable can lead to behaviors that hinder or prevent additional learning, e.g., avoiding additional challenges, losing interest in the topic after a setback, or limiting interactions and collaboration with peers \citep[e.g.,][]{Dweck1988,Dweck1999,Heine2001, Blackwell2007,Burnette2013,Smiley2016,Sanchez2020}. The potential benefits of a growth mindset and attendant mastery-oriented goals have been shown to be stronger in situations where students are more challenged \citep[e.g.,][]{Hoyert2008,Burnette2013}.  

In light of these findings, numerous interventions have been designed to help foster a growth mindset in students from school-age to university-level \citep[e.g.,][]{Blackwell2007,Paunesku2015,Lisberg2018,Beatty2019,Nallapothula2020,Chambers2022}. However, evidence of their effectiveness has been mixed; some studies show that encouraging a growth mindset improves academic achievement \citep[e.g.,][]{Aronson2002,Good2003,Yeager2011,Yeager2019} while others show much weaker or temporary benefits, in some cases only for certain groups of students \citep[e.g.,][]{Orosz2017,Sisk2018,Wang2021,Fink2018}. This heterogeneity was confirmed by a recent meta-analysis, and it was ascribed to the differing impact of growth mindset interventions depending on what population was being targeted, on how well the intervention was carried out, and on other contextual factors \citep{burnette2022}.

In addition, studies have found conflicting evidence about whether student mindsets change over time.  Among school-age children, some work has pointed to an increase in fixed mindset as children age, while other work has found the opposite \citep{Kinlaw2003,Gonida2006,Gunderson2017}.  At the university-level, \citet{Gunderson2017} found that the mindsets of high school and first- and second-year undergraduate students were similar, and \citet{Robins2002} found no change in mindset for students followed from high school until their fourth year at the university. 
On the other hand, a number of studies have found a drop in the growth mindset scores of students during introductory university-level courses in biology \citep{Dai2014}, computer science \citep{Scott2014,Flanigan2017}, and mathematics \citep{Shively2013}.  \citet{Malespina2022,Malespina2023} found that, after taking an undergraduate introductory physics course, students showed a decrease in their endorsement of a growth mindset, with the drop being more pronounced for female versus male students. Similarly, \citet{Limeri2020} found that students in a university organic chemistry course tended to shift towards a stronger fixed mindset and weaker growth mindset over the course of the semester, with a particularly strong trend for students who struggled in the course. The drop in growth mindset and increase in fixed mindset scores can be particularly pronounced when measured in a domain-specific fashion rather than simply relating to general intelligence \citep[e.g.,][]{Shively2013,Scott2014}.
Given this context, continued research is needed into how student mindsets depend on time and domain.

In this work,  
we present a study of the mindsets and attitudes of students in an introductory general education astronomy course at a public university in the southwestern United States.  We used standard mindset survey questions as well as a small subset of questions aimed at the behaviors and interpretations students demonstrated in the classroom and their ``in the moment'' mindset.  Along with investigating mindset measures, we also explored other attitudes that students held about learning astronomy — their affinity for the subject, their perceived competence, and the perceived degree of difficulty.  The arrival of the global COVID-19 pandemic and the associated shift to all-virtual learning during the course of the study provided an unexpected test of the stability of the attitudes and mindsets of astronomy undergraduate students over the course of five semesters.

In Section 2, we describe the survey instruments used in our study, and in Section 3, we review the resulting datasets.  We present our results in Section 4, and discuss the implications in Section 5.  We conclude in Section 6.

\section{Surveys\label{surveys}}

Each semester for five semesters, pre- and postsurveys were given to students taking an introductory general education astronomy course at a medium-sized public university in the southwestern United States. The course consists of twice-weekly lecture sessions covering basic principles of light and gravity, seasons and the Moon, stellar evolution, galaxy formation and evolution, and an introduction to cosmology.  Weekly laboratory sessions provide hands-on experience with related topics such as stellar parallax, optics, galaxy morphologies, and the discovery of the expanding universe. The course counts towards general education laboratory requirements and can be used to fulfill requirements for earning a minor in astronomy.  

We received approval from the university Institutional Review Board to conduct our surveys.
Students were informed that an optional, non-graded survey about student attitudes towards learning astronomy would be distributed at the beginning (presurvey) and end (postsurvey) of the semester. All participants in the surveys gave consent that they were over 18 years of age and that their responses would be reported in aggregate. In addition to the main survey questions, students were asked to report demographic information (year in school, gender, and race or ethnicity).  Final grades were collected at the end of the semester for students who completed the surveys.

For this exploratory study, we developed a baseline survey that included questions covering the following areas: affinity (whether students like astronomy), competence (whether students feel competent at astronomy), difficulty (whether students think astronomy is difficult), and mindset beliefs (the degree to which students subscribe to a growth versus a fixed mindset).  The questions on attitudes about learning astronomy (affinity, competence, difficulty) were included in order to provide context for student responses regarding their learning mindset in an astronomy class. They were adapted from an existing “Survey of Attitudes Toward Astronomy” by M. Zeilik \footnote{“Survey of Attitudes Toward Astronomy” by Michael Zeilik; http://www.flaguide.org/tools/attitude/astpr.htm}, which was modeled on a validated survey of students' attitudes toward introductory statistics developed by C. Schau \citep{Schau1995} and which has been used in past research on introductory astronomy students \citep{Zeilik1997,Zeilik1999}.   
Questions on mindset beliefs were modeled on standard questions used to assess learning mindsets, adapted from Dweck (1999) and from a related online “Mindset Quiz” by E. Diehl (private communication; \footnote{“Mindset Quiz” by Emily Diehl, Educator and Author; https://classroom20.com/forum/topics/motivating-students-with?commentId=649749\%3AComment\%3A1167061}), which has been used in a related undergraduate mindset intervention \citep{Hacisalihoglu2020}. 

After administering the baseline surveys over two semesters, there were indications that the mindset beliefs questions were not fully capturing students' approach to learning astronomy. For example, despite the fact that the responses to these standard mindset questions were overwhelmingly consistent with a growth mindset, in-person conversations with students and responses on student evaluations often included comments about being ``bad at math'' or requests to reduce the challenging material in the class in order to improve learning, statements that seemed to betray a classically fixed mindset approach.  To investigate this possibility, after two semesters, three sets of additional questions were added to investigate student mindsets in a more nuanced fashion. The first two sets touched on how well a student’s behaviors (Set 1 — mindset behaviors) or their interpretations of learning situations (Set 2 — mindset interpretations) aligned with a growth versus a fixed mindset.  The third set was aimed at taking students' growth mindset “temperature” by simply asking students where they would fall “right now” on a scale from 1-5, where 1 indicates they do not feel they have what it takes to succeed in astronomy, and 5 indicates they do (Set 3 — mindset temperature). During the final semester, a fourth set was added to assess whether a student felt they were generally strong at math/science and whether they felt that their success in the class depended more on their natural ability or on devoting sufficient time and effort to the task of learning (Set 4 — mindset self-assessment). As this study was exploratory in nature, future work, e.g., using student interviews, will be needed to further validate the survey questions.

Each survey question was associated with a question category, as shown in Tables~\ref{tab:affinity}-\ref{tab:mindsetselfassessment}.  Questions were scored using a Likert-scale (1–5). Responses to positively versus negatively-phrased questions were averaged separately to produce a category mean, standard deviation, and standard error of the mean. For attitude questions, this resulted in scores for positive affinity (liking astronomy) and negative affinity (disliking astronomy), positive competence (feeling competent at astronomy) and negative competence (not feeling competent at astronomy), and positive difficulty (viewing astronomy as difficult) and negative difficulty (not viewing astronomy as difficult).  In the case of the mindset questions, this resulted in separate scores for positive Mindset (growth mindset) and negative Mindset (fixed mindset), the suggested best practice \citep{Cook2017,Troche2020,Limeri2020}. 
In addition, a composite affinity, competence, difficulty, and Mindset score was generated after inverting the response scales for negatively-phrased questions, so that a score of 5 always corresponded to stronger affinity/competence/difficulty attitudes and a stronger growth mindset.  In this case, a score of 1 corresponded to weaker affinity/competence/difficulty attitudes and a weaker growth mindset. Again, question responses were averaged by category to produce a composite category mean, standard deviation, and standard error of the mean.

For classes taught in person by the main instructor, the concept of a growth versus fixed mindset was briefly mentioned, with general encouragement to adopt a growth mindset for improved learning.  These mentions happened during the first class period (after the presurvey was administered) and a few other times throughout the semester. No deliberate mention of learning mindsets was made in classes taught by alternate instructors or in classes taught using a distance education mode. However, as no significant change in learning mindset was detected between pre- and postsurvey responses in any case, we combine survey responses for all instructors and delivery modes in our analysis.

\section{Survey Datasets\label{datasets}}

Surveys were conducted at the beginning and end of each semester, but the unexpected start of the global COVID-19 pandemic in Spring 2020 
necessitated changes in both 
instruction delivery mode and 
survey administration. Details on each semester are described below; the resulting sample sizes are given in Table~\ref{tab:samples}.  

In Spring 2019, a single class was taught in person by the main instructor, a tenure-track faculty member.  Baseline pre- and postsurveys were administered on paper on the first and last days of class, with no extra credit incentive.  
During Fall 2019, two classes were taught in person by the same main instructor, a tenure-track faculty member, in back-to-back lecture periods.  Baseline pre- and postsurveys were administered on paper on the first and last days of class, with no extra credit incentive. 

In Spring 2020, two classes were taught by two alternate instructors, both tenure-track faculty members. The classes started in-person, but after the university transitioned online due to COVID-19 in March 2020, they were finished virtually (asynchronous online).  Baseline presurveys with additional questions from Sets 1, 2, and 3 were administered on paper, with no extra credit incentive.  Baseline postsurveys with additional questions from Sets 1, 2, and 3 were administered online via Qualtrics \footnote{www.qualtrics.com}, with no extra credit incentive. 

Due to the continuation of the COVID-19 pandemic during Fall 2020, a single class was taught using a distance education mode of delivery (asynchronous online). The course was taught by an alternate instructor, a tenure-track faculty member.  Baseline pre- and postsurveys with additional questions from Sets 1, 2, and 3 were administered online via Qualtrics, with no extra credit incentive. 

In Spring 2021, due to the continuation of the COVID-19 pandemic, two classes were taught using a distance education mode of delivery (asynchronous online). The courses were taught by the main instructor and an alternate instructor, both tenure-track faculty members.  Baseline pre- and postsurveys with additional questions from Sets 1, 2, 3, and 4 were administered online via Qualtrics, with an extra credit incentive. 

\section{Results\label{results}}

\subsection{Semester-to-Semester Variation}
\label{sec:overall}

\begin{figure*}
\centering
	\includegraphics[width=1.0\textwidth]{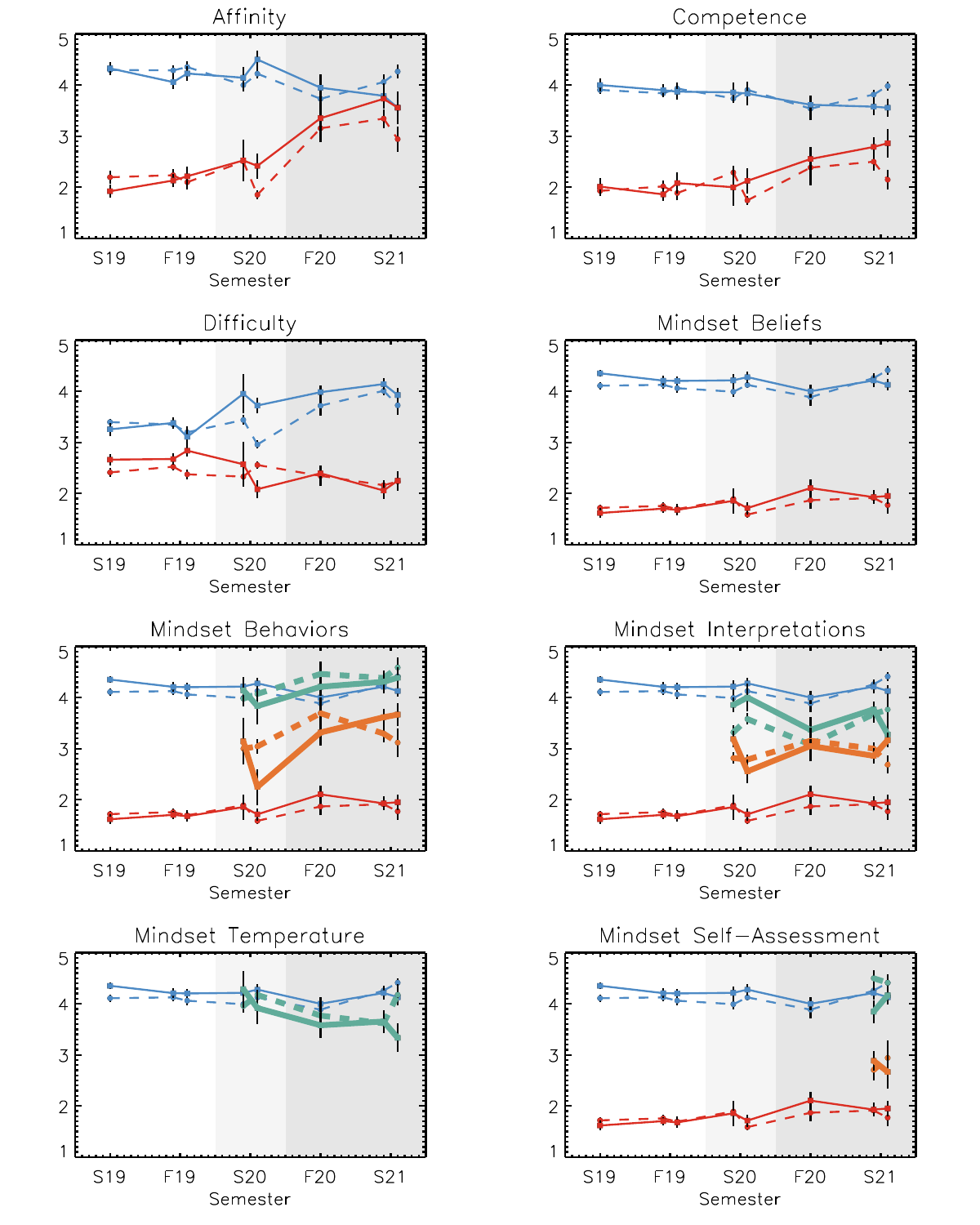} 
\caption{Mean presurvey (circles, dashed lines) and postsurvey (squares, solid lines) scores versus semester. Error bars represent the standard error of the mean.  Background shading indicates the teaching modality for each semester, whether in-person (white), hybrid (light grey), or fully online (dark grey).  In the top four panels, positive question (blue lines) and negative question (red lines) scores are shown, where applicable.  In the bottom four panels, results from positive (thick green lines) and negative (thick orange lines) alternative Mindset questions are compared with the standard Mindset Belief questions (blue and red thin lines).}\label{fig:surveyresults}
\end{figure*}

First, we investigated the stability of student attitudes and mindsets about learning astronomy over the course of five semesters straddling the beginning of the COVID-19 pandemic.
Figure~\ref{fig:surveyresults} shows the survey results as a function of semester (Spring 2019—Spring 2021).  The top four panels show the mean affinity, competence, difficulty, and mindset beliefs scores with error bars representing the standard error of the mean. During the Spring and Fall 2019 semesters, students overall showed an affinity for astronomy (high positive affinity, low negative affinity) and felt that they were relatively competent at astronomy (high positive competence, low negative competence). They perceived astronomy to have an intermediate level of difficulty (intermediate positive difficulty, intermediate negative difficulty). They agreed strongly with statements 
indicating a growth mindset, i.e., high positive mindset beliefs and low negative mindset beliefs. 

The semester when COVID-19 began (Spring 2020) $-$ causing
in-person courses to be switched to online delivery $-$ 
is indicated with light grey shading, and the two semesters taught entirely in an asynchronous online mode due to COVID-19 (Fall 2020 and Spring 2021) are shown with dark grey shading.  While scores are relatively stable prior to COVID-19 and during the hybrid Spring 2020 semester, during the subsequent 2020/2021 year of COVID-19, the responses even in the presurvey appear to drop in positive 
affinity and increase strongly in negative affinity.  The same trends appear more subtly in competence.  For difficulty, the reverse trend appears, with positive 
difficulty scores increasing and negative difficulty scores decreasing somewhat during COVID-19. For Mindset, there is a hint that positive Mindset scores were slightly lower and negative Mindset scores slightly higher after COVID-19 began, but the change was largely within the uncertainties.  Thus, it appears that mindset beliefs are a particularly stable aspect of students' mental landscape.

\subsection{Alternative Mindset Questions}\label{sec:altmindset}
To understand whether students' stated beliefs were accompanied by growth-oriented behaviors and interpretations of classroom experiences, we explored the responses to more nuanced mindset questions and how they compared to our measurements of students' mindset beliefs, again, as a function of semester. In each of the bottom four panels, we reproduce the mean mindset beliefs scores and take these to represent the degree to which students’ broad beliefs align with a growth versus a fixed Mindset, respectively.  We then overplot the scores for our alternative Mindset questions — those touching on mindset behaviors (Set 1), mindset interpretations (Set 2), Mindset “Temperature”, i.e., a student’s mindset ``in the moment'' (Set 3), and a mindset self-assessment of what factors (innate talent versus time and effort) are most important for success (Set 4).  While we have data from fewer semesters for these question sets, we can begin to see some interesting trends.  In terms of mindset behaviors, while students' positive Mindset scores are similar to their mindset beliefs, they show significantly higher negative Mindset scores. For mindset interpretations, both the positive Mindset scores are lower and the negative Mindset scores are higher compared to students’ mindset beliefs.  The mindset temperature question shows slightly lower positive Mindset scores as compared to students’ positive mindset beliefs scores.  Finally, in terms of mindset self-assessment, students showed similar positive Mindset scores but significantly higher negative Mindset scores compared with their mindset beliefs.  

Overall, from the offsets seen between students’ mindset beliefs and their responses to more nuanced Mindset questions, it appears that alternate questions can be used to probe less stable characteristics of students' approach to the learning. It may also suggest that while students have incorporated aspects of a growth mindset into their stated worldview, they do not always act or respond accordingly in the learning environment.

\subsection{Stability of Survey Responses Over the Semester as a Function of Demographics}
\label{sec:prepost}

We also investigated the stability of students' attitudes and mindsets over an individual semester. 
Given our limited sample sizes, we restrict our analysis to exploring the correlations between the presurvey versus postsurvey scores. We report the associated Spearman’s correlation coefficients, $\rho$, in Table~\ref{tab:rho}, with statistically significant correlations ($p<0.05$) indicated in boldface. Results are given for all students and also as a function of demographics. For simplicity, in this analysis we consider only composite mean scores (i.e., combining positively and negatively phrased questions, as described in Section~\ref{surveys}).

Overall, we find that the pre- and postsurvey responses are largely consistent with one another for the affinity, competence, difficulty, and mindset beliefs scores, showing moderate Spearman’s correlation coefficients of $\rho$ = 0.5–0.6 when considering all students responses in aggregate.  The results are similar when dividing the students by age (``younger'', 1st year only, versus ``older'', $>$1st year), gender (``female'' versus ``male''), and race or ethnicity (``Hispanic'' versus ``White'').  
We note that students who selected an alternate gender identity ($N=4$) or an alternate race or ethnicity category ($N=24$) were not numerous enough for separate quantitative analysis.

The full sample also shows moderate correlations ($\rho$ = 0.5-0.6) between pre- and postsurvey mindset interpretations and mindset temperature scores, with slightly stronger correlations in both cases for ``younger'' ($\rho$ = 0.7) versus ``older'' ($\rho$ = 0.4) students.
In addition, mindset temperature scores were slightly more  correlated for ``male'' ($\rho$ = 0.7) versus ``female'' ($\rho$ = 0.5) students. A weak correlation ($\rho$ = 0.3–0.4) is found between 
the pre- and postsurvey mindset behaviors scores, with slightly stronger correlations for ``White'' ($\rho$ = 0.5) versus ``Hispanic'' ($\rho$ = 0.4) students.  A similarly weak correlation ($\rho$ = 0.3–0.4) is seen between the pre- 
and postsurvey mindset self-assessment scores.

Thus, students are quite consistent in their reported attitudes (affinity, competence, difficulty) and mindset beliefs across the span of a given semester, although there may be some slight variation in the responses to the alternative Mindset questions as a function of demographic groups.

\subsection{Relationship of Post-Survey Responses to Final Grades as a Function of Demographics}
\label{sec:postfinal}

Given the potential for a student's performance in the class to influence their attitudes about learning astronomy, we investigated whether students' attitudes and mindsets at the end of the semester were correlated with their final grade percentage in the class. Again, given our limited sample sizes, we restrict our analysis to correlations, reporting the associated Spearman’s correlation coefficients, $\rho$, in Table~\ref{tab:rho}, with statistically significant values ($p<0.05$) indicated in boldface. Results are given for all students and also as a function of demographics. For simplicity, in this analysis we again consider only composite mean scores (i.e., combining positively and negatively phrased questions, as described in Section~\ref{surveys}).

Overall, we find that the postsurvey responses and final grades are weakly correlated for affinity ($\rho$ = 0.3), including when dividing by age, but with hints of weaker correlations for male versus female and for White versus Hispanic students.
We find a moderate correlation for competence ($\rho$ = 0.4), including when dividing by gender, 
but with a stronger correlation for ``younger'' ($\rho$ = 0.5) versus ``older'' ($\rho$ = 0.3) students. The competence scores also hint at a much weaker correlation for White versus Hispanic students. Overall, mindset beliefs scores are only weakly correlated with final grades ($\rho$ = 0.2), with even weaker correlations suggested for male versus female and White versus Hispanic students.
mindset interpretations scores were moderately correlated ($\rho$ = 0.4) only for ``older'' versus ``younger'' students. mindset temperature scores appear moderately correlated ($\rho$ = 0.4) overall, 
but with hints of weaker correlations for male versus female and White versus Hispanic students.

Thus, despite the potential for a student's course grade to influence their attitudes about learning astronomy, the results from our correlation analysis are somewhat mixed, with hints of variation between demographic groups.  These data suggest that the attitudes of White and male students, in particular, may be less influenced by their final grade in the class than other demographic groups.
In addition, the stronger correlations between affinity, competence, and mindset temperature responses and final grades as compared to the results for mindset beliefs scores are consistent with the idea that mindset beliefs questions are tracking a more stable characteristic than that probed by our attitude and nuanced mindset questions and one that is less influenced by the grade in a particular class.

\section{Discussion\label{discussion}}

Our main finding is that university students in introductory astronomy classes broadly agree with general statements supporting a growth mindset perspective.  This is true for both Fall and Spring semester cohorts, as well as for different course delivery modes (in-person, hybrid in-person and virtual, and fully virtual).  The unexpected arrival of the COVID-19 pandemic in 2020, partway through our study, brought with it a host of unprecedented challenges that affected students striving to continue their education in myriad ways.
Indeed, we see evidence that this period took a toll 
on students' attitudes about learning astronomy (and presumably other topics).
Whether it was due to overall heightened stress levels during this time or due to the sudden change in course delivery mode (from being in-person to taking only online classes), students began the Fall 2020 and Spring 2021 semesters expressing significantly lower affinity for astronomy, somewhat lower levels of perceived competence, and slightly higher levels of perceived difficulty than during the previous few semesters. Yet this same time period seemed to have a negligible impact on students’ learning mindset, as measured using standard growth mindset questions.

The degree to which students agree with growth mindset statements also did not change substantially from the beginning to the end of a given semester. Minimal correlations were found between 
postsurvey responses to standard mindset questions versus the final grade in the class, emphasizing the durability of students’ belief in a growth mindset, but also failing to corroborate previous findings of a positive correlation between a growth mindset and academic achievement at the university level \citep{Limeri2020}. The other attitude measures we tracked — the affinity students have for astronomy, the perceived competence at astronomy, and the perceived difficulty of astronomy — all showed more variation semester-to-semester than the learning mindset scores, showing their learning mindset beliefs to be a remarkably stable feature of students’ mental landscape.

\begin{figure*}
\centering
	\includegraphics[width=0.5\textwidth]{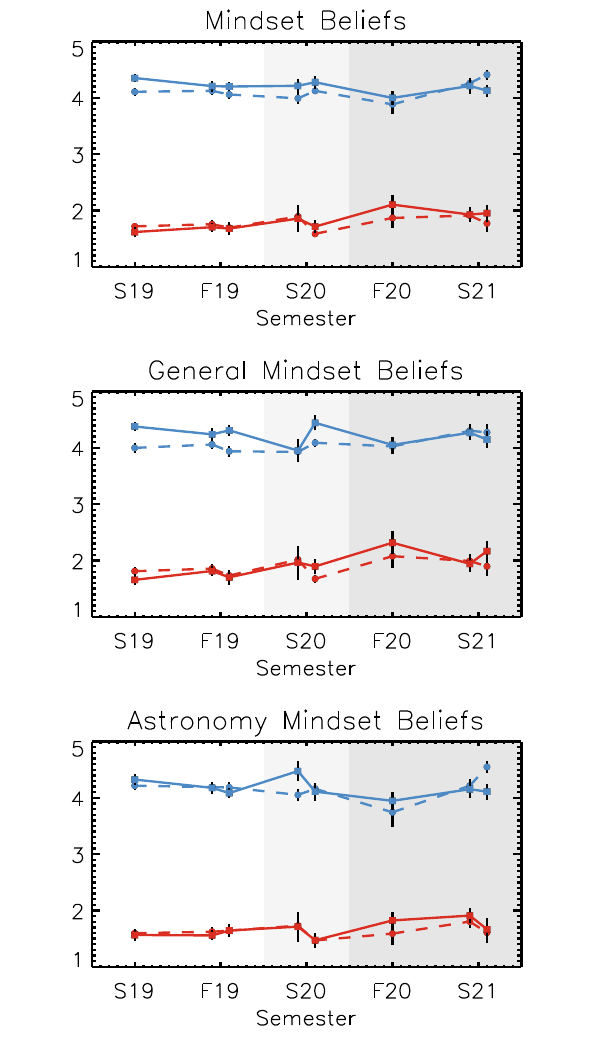} 
\caption{
Mean presurvey (circles, dashed lines) and postsurvey (squares, solid lines) scores versus semester for mindset beliefs questions. All mindset belief questions are shown in the top panel, as in Figure~\ref{fig:surveyresults}, while the lower two panels show the results for general mindset belief questions (middle panel) versus astronomy-specific mindset belief questions (bottom panel), as discussed in the text.
Positive question (blue) and negative question (red) 
scores are shown.  Error bars represent the standard error of the mean.  Shading indicates the teaching modality for each semester, whether in-person (white), hybrid (light grey), or fully online (dark grey)}.\label{fig:generalVastro}
\end{figure*}

These findings are particularly interesting in the context of previous work showing that students in a range of other undergraduate STEM courses (i.e., biology, computer science, math, organic chemistry, physics) tend to show a shift towards fixed mindset beliefs over the course of the semester \citep[e.g.,][]{Dai2014, Flanigan2017, Scott2014, Shively2013, Limeri2020, Malespina2022, Malespina2023}. 
Studies have shown that students’ mindset can depend on the relevant academic domain, with domain-specific mindsets evolving more significantly towards a fixed mindset than students' general mindset over the course of a semester \citep{Shively2013,Scott2014,Gunderson2017}. However, we found that the results of our study were similar (Figure~\ref{fig:generalVastro}) when we considered only those Mindset questions referring to general intelligence versus those referring to astronomy-specific intelligence (see Table~\ref{tab:mindsetbeliefs}). The reason for this difference between 
students taking astronomy classes versus 
those taking other STEM subjects is not clear. 
It is possible that undergraduate student mindsets depend on the particular make-up of the student cohort, e.g., on the quality of student motivation or engagement in the class \citep{Flanigan2017}, or whether a course is taken as a required (higher stakes) prerequisite for entering an undergraduate major versus as fulfillment of a (lower stakes) general education requirement. The mindset students bring might also be influenced by the extent to which success in a given field is thought to require ``brilliance'' versus hard work; for example, astronomy has been shown to be somewhat less strongly associated with the brilliance narrative than fields like math, physics, and computer science, although it is comparable to fields like biology and chemistry \citep{Leslie2015}. Alternately, it is possible that there has been a gradual evolution in student mindsets over the past two decades, as the concept of a growth mindset has become more widely popularized.  Future study will be needed to better understand these findings.

The picture becomes more nuanced when we look at other measures of students’ learning mindset. When asked about behaviors that are aligned or anti-aligned with a growth mindset — e.g., asking for help or looking up additional resources, focusing on what came easily or avoiding what was challenging — student fixed mindset scores are significantly higher than when their mindset is assessed using broad statements of what they believe.  Similarly, when asked about how they interpret situations in the class — e.g., easily understanding what the teacher says reflecting that they already know enough, saying learning is improved when a class does not focus on areas where they are weak versus saying their learning is improved when a class covers challenging topics, or interpreting having to work hard to understand something as an indication they are not good at that topic — student growth (fixed) Mindset scores are significantly lower (higher) than the scores for overall beliefs.  When simply asked whether their success in the class would depend mostly on their natural math or science ability versus on the time and effort they spend, 
student fixed mindset scores are somewhat higher.  Interestingly, when asked to put themselves on a scale from 1-5, where 1 indicated they do not feel they have what it takes to succeed, and 5 indicated that they do, the drop in growth mindset score was minimal during the hybrid Spring 2020 semester, but seemed to become more pronounced in the fully-virtual classrooms.  Taken together, all of these different approaches suggest that while introductory astronomy students say they very much ``believe'' in a growth mindset — 
that intelligence is malleable and that you can get smarter by putting in time and effort — they do not necessarily act or respond that way consistently in the class setting.  For example, they do not necessarily engage in the behaviors that would would allow them to “get smarter with time and effort”, or they do not necessarily lean into that next challenge that would help them grow. Instead, they may run away from challenge or interpret struggle as a sign that they are not good at something rather than simply as an indication that they do not understand that topic ``yet'' \citep{Dweck2006}.  This discrepancy between stated beliefs and classroom behaviors or interpretations may indicate that standard mindset questions do not fully capture student learning mindsets, at least in the undergraduate classroom setting.  Alternately, once again, it is possible that this discrepancy reflects evolution over the past two decades. Having grown up with the idea of a growth mindset, students may have come to adopt a growth mindset into their stated beliefs, while not necessarily having the skills or support needed to put a growth mindset into practice.

We note that there is precedence in the literature for 
the finding that individuals can hold aspects of a 
growth mindset and a fixed mindset at the same time, 
both among students \citep[e.g.,][]{Kalendar2022,Malespina2022,Malespina2023} as well as among faculty serving on graduate admissions committees \citep[e.g.,][]{Scherr2017}.
Indeed, it was this complexity that motivated the recommendation to track growth and fixed mindset scores separately \citep[e.g.,][]{Cook2017,Troche2020,Limeri2020}. 
As a specific example, \citet{Little2019} conducted end-of-semester interviews of students in an introductory physics class and found evidence that some students would express aspects of a fixed mindset and yet would demonstrate a high level of hard work and persistence, i.e., behavior that is more aligned with a growth mindset. The apparent contrast with what we found (growth mindset beliefs yet fixed mindset behaviors/interpretations) likely stems from differences in the student population studied: \citet{Little2019} chose to interview STEM majors who were likely motivated to earn a high grade in a second semester introductory physics class, whereas students taking the introductory astronomy courses we studied were drawn from a wide range of majors and were most often taking the course simply to fulfill a general education lab requirement.

No strong differences in the pre/post correlations for affinity, competence, difficulty, and mindset beliefs were found as a function of student demographics.  This suggests that our findings, particularly on the durability of mindset beliefs, are likely to be fairly representative of the student body as a whole.  While the smaller sample size makes it difficult to draw firm conclusions on the more nuanced Mindset questions, it appears that in some cases there is less pre/post correlation and 
more variation between demographic groups, suggesting that these questions may be probing less stable characteristics that are more open to influence by experiences in the classroom.

A longstanding goal has been to design interventions that encourage students to adopt a growth mindset  in hopes of ultimately improving their academic performance and persistence.  From this perspective, our results are encouraging.  The fact that university introductory astronomy students are in strong agreement with growth mindset beliefs, i.e., with the idea that they can get smarter given time and effort, is a great starting point for improving their learning.  However, interventions designed to foster a growth mindset may encounter the ceiling effect that we saw in this work, when the level of growth mindset agreement is already high \citep[e.g.,][]{Beatty2019}. More importantly, 
there is clearly more work to do in terms of helping students understand what good learning actually feels like and helping them take a growth mindset to heart in terms of their learning behaviors and interpretations.  Many things in the classroom setting can “trigger” a fixed mindset. For example, students will inevitably compare themselves to other students or worry they do not measure up in some regard \citep[e.g.,][]{Little2019}; they will encounter a setback or have to struggle to understand the material.  In addition, some students may encounter financial or personal challenges that make it difficult to follow through on devoting the time and effort needed to succeed academically. 
To maximize their ability to learn, therefore, students need to not only believe a growth mindset is a good thing in the abstract, but they must also develop specific skills for responding mindfully to classroom experiences and must have sufficient time and resources to devote to their education.
This suggests that the most useful interventions will be those that focus explicitly on encouraging and supporting specific growth-oriented habits and behaviors that have been demonstrated to improve learning rather than on influencing overarching beliefs.
\\

\section{Conclusions\label{conclusions}}

In order to understand the attitudes and mindsets students have towards learning astronomy, we conducted pre- and postsurveys in an introductory general education astronomy course taught at a medium-sized public university in the southwestern United States. Students showed strong and consistent agreement with standard mindset questions, both over the span of a single semester as well as across semesters coinciding with the beginning of the COVID-19 pandemic and the associated, rapid shift to all-virtual learning.  Measures of other learning attitudes showed more variation over the course of the COVID-19 pandemic than students’ expressed mindset beliefs, emphasizing that students appear to strongly believe in their ability to grow their intelligence through time and effort.  Responses to these attitude and mindset questions did not depend strongly on student demongraphics, suggesting that these results are representative of the student population as a whole.  However, on more nuanced measures of students’ learning mindset — i.e., questions that probe the behaviors and interpretations students demonstrate in the classroom, their ``in the moment'' mindset, and students’ self-assessment of what factors are most important for their success — students showed lower (higher) alignment with a growth (fixed) mindset.  
This suggests that standard mindset survey questions may not (or may no longer) capture a complete measure of student mindsets in the undergraduate classroom environment. Specifically, 
while students may have adopted the tenets of a growth mindset into their stated beliefs, they do not necessarily act or respond accordingly when faced with the inevitable challenges of learning something new.  
Interventions aimed at improving student learning though influencing learning mindsets should therefore consider focusing less on convincing students to adopt a growth mindset in the abstract and 
more on training students to become mindful of their responses to classroom experiences $-$ i.e., understanding what real learning feels like, so they can interpret academic challenges constructively $-$ and supporting them in adopting behaviors that foster rather than undermine their learning.

%


\begin{table*}[bt!]
\caption{Baseline Survey $-$ Astronomy affinity}\label{tab:affinity}
\begin{tabular}{l l r}
\toprule
\bf{Question (positive)} & &  \bf{Source}\\
\hline
	6. I will like (liked) astronomy.   & \textemdash& Zeilik [36] (adapted) \\
	16. I will enjoy (enjoyed) taking this astronomy course. & \textemdash  & Zeilik [36] \\
\hline
\bf{Question (negative)} &  & \bf{Source} \\
\hline
	10. I will be (was) under stress during astronomy class. & \textemdash & Zeilik [36] \\
	15. I will feel (felt) insecure when I have (had) to do astronomy homework. & \textemdash & Zeilik [36] \\
	22. The thought of taking an (another) astronomy course scares me. & \textemdash & Zeilik [36] (adapted) \\
\hline
\end{tabular}

\end{table*}

\begin{table*}[bt!]
\caption{Baseline Survey $-$ Astronomy perceived competence}\label{tab:competence}
\begin{tabular}{l l r}
\toprule
\bf{Question (positive)} & & \bf{Source}\\
\hline
	14. I will find (found) it easy to understand astronomy concepts.   & \textemdash  & Zeilik [36] (adapted) \\
20. I can learn astronomy. & \textemdash& Zeilik [36] \\
	21. I will understand (understood) how to apply analytical reasoning to astronomy. & \textemdash& Zeilik [36] \\
\hline
\bf{Question (negative)} & & \bf{Source} \\
\hline
	2. I will have (had) trouble understanding astronomy because of how I think. & \textemdash & Zeilik [36] \\
	9. I will have (had) no idea of what's going on in astronomy. & \textemdash & Zeilik [36] \\
\hline
\end{tabular}
\end{table*}

\begin{table*}[bt!]
\caption{Baseline Survey $-$ Astronomy perceived difficulty}\label{tab:difficulty}
\begin{tabular}{l l r}
\toprule
\bf{Question (positive)} & & \bf{Source}\\
\hline
8. Learning astronomy requires a great deal of hard work.   & \textemdash & Zeilik [36] (adapted) \\
19. Astronomy is a complicated subject. & \textemdash & Zeilik [36] (adapted) \\
26. Astronomy concepts are hard for most people to understand. & \textemdash & Zeilik [36] (adapted) \\
\hline
\bf{Question (negative)} & & \bf{Source} \\
\hline
1. Astronomy is a subject learned quickly by most people. & \textemdash & Zeilik [36] \\
13. Astronomy is not highly technical. & \textemdash & Zeilik [36] (adapted) \\
\hline
\end{tabular}

\end{table*}

\begin{table*}[bt!]
\caption{Baseline Survey $-$ Mindset beliefs. Note that $^{*}$ are general mindset questions and $^{**}$ are astronomy-specific mindset questions, as discussed in the text. }\label{tab:mindsetbeliefs}
\begin{tabular}{l l r}
\toprule
\bf{Question (positive)} & & \bf{Source}\\
\hline
3. No matter how much intelligence you have, you can always change it quite a bit.$^{*}$ & \textemdash & \citet{Dweck1999} \\
12. I appreciate when people give me feedback about my performance.$^{*}$ & \textemdash & Diehl [40] \\
	17. I will be (was) able to learn how to apply concepts in astronomy.$^{**}$ & \textemdash & This work \\
24. You can substantially change how intelligent you are.$^{*}$ & \textemdash & \citet{Dweck1999} \\
27. The harder you work at something like astronomy, the better you will be at it.$^{**}$ & \textemdash & Diehl [40] \\
28. An important reason why I am studying astronomy is that I like to learn new things.$^{**}$ & \textemdash & Diehl [40] \\
\hline
\bf{Question (negative)} & & \bf{Source} \\
\hline
4. I often feel upset or angry when I get feedback about my performance.$^{*}$ & \textemdash & Diehl [40] \\
5. Only a few people will be truly good at astronomy $—$ you have to be “born with it.”$^{**}$ & \textemdash & Diehl [40] \\
7. You can learn new things, but you can’t really change how intelligent you are.$^{*}$& \textemdash & \citet{Dweck1999} \\
11. Truly smart people do not need to struggle to learn things like astronomy.$^{**}$ & \textemdash & Diehl [40] \\
18. Your intelligence is something very basic about you that you can’t change very much.$^{*}$ & \textemdash & \citet{Dweck1999} \\
23. Astronomy is much easier for people of a certain gender or from a certain culture.$^{**}$ & \textemdash & Diehl [40] \\
25. Learning new things is stressful for me, and I tend to avoid doing so.$^{*}$ & \textemdash & Diehl [40] \\
\hline
\end{tabular}

\end{table*}

\begin{table*}[bt!]
\caption{Additional Question Set 1 $-$ Mindset behaviors}\label{tab:mindsetbehaviors}
\begin{tabular}{l l r}
\toprule
\bf{Question (positive)} &  &\bf{Source}\\
\hline
	34. When I am (was) confused about something in astronomy, I will look (looked) up & \textemdash & This work\\
	additional resources or get (got) help from teachers or classmates until I understand. &  & \\
\hline
\bf{Question (negative)} & & \bf{Source} \\
\hline
	29. When I am (was) having difficulty understanding a particular topic in astronomy, & \textemdash & This work \\
	I will try (tried) to focus my energy instead on the areas where I feel stronger. &  & \\
\hline
\end{tabular}

\end{table*}

\begin{table*}[bt!]
\caption{Additional Question Set 2 $-$ Mindset interpretations
}\label{tab:mindsetinterpretations}
\begin{tabular}{l l r}
\toprule
\bf{Question (positive)} & & \bf{Source}\\
\hline
32. I learn the most when a class covers topics or skills that I find challenging. & \textemdash & This work \\
\hline
\bf{Question (negative)} & & \bf{Source} \\
\hline
30. My learning is improved when a class does not focus too much on topics & \textemdash & This work \\
or skills that I am not good at. &  &  \\
	31. If I quickly and easily understand (understood) what the teacher says (said), & \textemdash & This work\\ 
	I am (was) satisfied I know (knew) enough about that topic. &  &  \\
	33. When I have (had) to work hard to understand a topic in astronomy, this & \textemdash & This work \\
	means (meant) that I am (was) not good at that topic. &  &  \\
\hline
\end{tabular}

\end{table*}

\begin{table*}[bt!]
\caption{Additional Question Set 3 $-$ Mindset temperature
}\label{tab:mindsettemperature}
\begin{tabular}{l l r}
\toprule
	\bf{Question} & & \bf{Source}\\
\hline
	39. Where are you right now on the following scale?: & \textemdash & This work \\
	\includegraphics[width=0.6\textwidth]{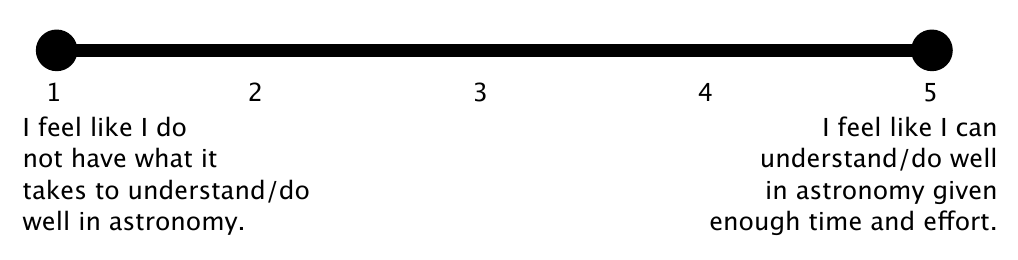}
	& &  \\
\hline
\end{tabular}

\end{table*}

\begin{table*}[bt!]
\caption{Additional Question Set 4 $-$ Mindset self-assessment}
\label{tab:mindsetselfassessment}
\begin{tabular}{l l r}
\toprule
\bf{Question (positive)} & & \bf{Source}\\
\hline
	38. The primary factor determining my level of success in astronomy will be (was) & \textemdash & This work\\
my ability to devote sufficient time and effort. & & \\
\hline
\bf{Question (negative)} & & \bf{Source} \\
\hline
	36. The primary factor determining my level of success in astronomy will be (was) & \textemdash & This work\\
my natural math or science abilities. & & \\
\hline
\bf{Additional Questions} & & \bf{Source} \\
\hline
35. I am naturally strong in terms of my math or science abilities. & \textemdash & This work \\
	37. I will be (was) able to devote the time and effort needed to learn astronomy. & \textemdash & This work \\
\hline
\end{tabular}

\end{table*}

\begin{table*}[bt!]
\caption{Survey Samples. Note that $^{*}$ indicates the number of students enrolled at the start of the semester.}\label{tab:samples}
\begin{tabular}{lccccc}
\toprule
Semester & N$_{Pre}$ & N$_{Post}$ & N$_{Pre+Post}$ & N$_{Enrolled}$$^{*}$ & Included Questions \\
\hline
Spring 2019 & 63 & 43 & 39 & 67 & Baseline \\
Fall 2019 (2 classes) & 61+43 & 44+31 & 41+26 & 70+52 & Baseline \\
Spring 2020 (2 classes) & 47+70 & 7+12 & 6+11 & 52+80 & Baseline, Sets 1-3\\
Fall 2020 & 13 & 19 & 4 & 154 & Baseline, Sets 1-3 \\
Spring 2021 (2 classes) & 17+34 & 17+27 & 11+24 & 46+45 & Baseline, Sets 1-4 \\
\hline
\end{tabular}

\end{table*}

\begin{table*}[bt!]
\caption{Spearman's $\rho$ Correlations$^{*}$ - Age, Gender, Race/Ethnicity. 
	Statistically significant correlations ($p<0.05$) are shown in bold.}\label{tab:rho}
\begin{tabular}{lccccccc}
\toprule
Presurvey vs. Postsurvey & All & Younger & Older & Female & Male & Hispanic & White \\
\hline
                         Affinity & \bf{  0.543  } & \bf{  0.487  } & \bf{  0.596  } & \bf{  0.525  } & \bf{  0.526  } & \bf{  0.595  } & \bf{  0.549      } \\
                       Competence & \bf{  0.592  } & \bf{  0.573  } & \bf{  0.601  } & \bf{  0.531  } & \bf{  0.609  } & \bf{  0.572  } & \bf{  0.600      } \\
                       Difficulty & \bf{  0.565  } & \bf{  0.460  } & \bf{  0.627  } & \bf{  0.469  } & \bf{  0.561  } & \bf{  0.548  } & \bf{  0.630      } \\
                  Mindset Beliefs & \bf{  0.640  } & \bf{  0.692  } & \bf{  0.608  } & \bf{  0.693  } & \bf{  0.564  } & \bf{  0.685  } & \bf{  0.542      } \\
                Mindset Behaviors & \bf{  0.295  } & \bf{  0.486      } &   0.283    & \bf{  0.320      } &   0.492    & \bf{  0.418  } & \bf{  0.496      } \\
          Mindset Interpretations & \bf{  0.559  } & \bf{  0.696  } & \bf{  0.427  } & \bf{  0.512      } &   0.577    & \bf{  0.534  } & \bf{  0.520      } \\
              Mindset Temperature & \bf{  0.515  } & \bf{  0.694  } & \bf{  0.387  } & \bf{  0.488  } & \bf{  0.683  } & \bf{  0.629      } &   0.293        \\
          Mindset Self-Assessment & \bf{  0.375      } &   0.220    & \bf{  0.449      } &   0.364        &   0.147    & \bf{  0.484      } &  -0.099        \\
\toprule
Post-Survey vs. Final Score & All & Younger & Older & Female & Male & Hispanic & White \\
\hline
                         Affinity & \bf{  0.257  } & \bf{  0.325  } & \bf{  0.218  } & \bf{  0.307      } &   0.108    & \bf{  0.346      } &  -0.046        \\
                       Competence & \bf{  0.399  } & \bf{  0.504  } & \bf{  0.335  } & \bf{  0.400  } & \bf{  0.356  } & \bf{  0.526      } &   0.127        \\
                           Difficulty &  -0.011        &  -0.056        &   0.022        &   0.053        &  -0.119        &  -0.013        &  -0.020        \\
                  Mindset Beliefs & \bf{  0.187      } &   0.219        &   0.169    & \bf{  0.227      } &   0.022    & \bf{  0.271      } &  -0.097        \\
                    Mindset Behaviors &   0.142        &   0.044        &   0.286        &   0.217        &  -0.586        &   0.147        &   0.086        \\
              Mindset Interpretations &   0.230        &   0.067    & \bf{  0.414      } &   0.289        &  -0.222        &   0.288        &   0.072        \\
              Mindset Temperature & \bf{  0.424      } &   0.386    & \bf{  0.476  } & \bf{  0.497      } &  -0.112    & \bf{  0.530      } &   0.012        \\
              Mindset Self-Assessment &   0.221        &   0.278        &   0.203    & \bf{  0.520      } &  -0.109        &   0.283        &  -0.351        \\
\hline
\end{tabular}
 
\end{table*}


\begin{acknowledgments}
The authors would like to thank Kristian Finlator for helpful discussions and Jason Jackiewicz, Kristian Finlator, and Joseph Burchett for allowing us to collect survey responses in their introductory astronomy classrooms. Figures 1 and 2 were created using color-blind-friendly IDL color tables developed by Paul Tol (https://personal.sron.nl/pault/).

M.K.M.P. acknowledges support from NSF grant AAG-1813016.

\end{acknowledgments}
\clearpage

\begin{thebibliography}{48}%
\makeatletter
\providecommand \@ifxundefined [1]{%
 \@ifx{#1\undefined}
}%
\providecommand \@ifnum [1]{%
 \ifnum #1\expandafter \@firstoftwo
 \else \expandafter \@secondoftwo
 \fi
}%
\providecommand \@ifx [1]{%
 \ifx #1\expandafter \@firstoftwo
 \else \expandafter \@secondoftwo
 \fi
}%
\providecommand \natexlab [1]{#1}%
\providecommand \enquote  [1]{``#1''}%
\providecommand \bibnamefont  [1]{#1}%
\providecommand \bibfnamefont [1]{#1}%
\providecommand \citenamefont [1]{#1}%
\providecommand \href@noop [0]{\@secondoftwo}%
\providecommand \href [0]{\begingroup \@sanitize@url \@href}%
\providecommand \@href[1]{\@@startlink{#1}\@@href}%
\providecommand \@@href[1]{\endgroup#1\@@endlink}%
\providecommand \@sanitize@url [0]{\catcode `\\12\catcode `\$12\catcode `\&12\catcode `\#12\catcode `\^12\catcode `\_12\catcode `\%12\relax}%
\providecommand \@@startlink[1]{}%
\providecommand \@@endlink[0]{}%
\providecommand \url  [0]{\begingroup\@sanitize@url \@url }%
\providecommand \@url [1]{\endgroup\@href {#1}{\urlprefix }}%
\providecommand \urlprefix  [0]{URL }%
\providecommand \Eprint [0]{\href }%
\providecommand \doibase [0]{https://doi.org/}%
\providecommand \selectlanguage [0]{\@gobble}%
\providecommand \bibinfo  [0]{\@secondoftwo}%
\providecommand \bibfield  [0]{\@secondoftwo}%
\providecommand \translation [1]{[#1]}%
\providecommand \BibitemOpen [0]{}%
\providecommand \bibitemStop [0]{}%
\providecommand \bibitemNoStop [0]{.\EOS\space}%
\providecommand \EOS [0]{\spacefactor3000\relax}%
\providecommand \BibitemShut  [1]{\csname bibitem#1\endcsname}%
\let\auto@bib@innerbib\@empty
\bibitem [{\citenamefont {Dweck}\ and\ \citenamefont {Leggett}(1988)}]{Dweck1988}%
  \BibitemOpen
  \bibfield  {author} {\bibinfo {author} {\bibfnamefont {C.~S.}\ \bibnamefont {Dweck}}\ and\ \bibinfo {author} {\bibfnamefont {E.~L.}\ \bibnamefont {Leggett}},\ }\bibfield  {title} {\bibinfo {title} {{A social{\^{}}cognitive approach to motivation and personality.}},\ }\href {https://doi.org/10.1037/0033-295X.95.2.256} {\bibfield  {journal} {\bibinfo  {journal} {Psychological Review}\ }\textbf {\bibinfo {volume} {95}},\ \bibinfo {pages} {256} (\bibinfo {year} {1988})}\BibitemShut {NoStop}%
\bibitem [{\citenamefont {Dweck}(1999)}]{Dweck1999}%
  \BibitemOpen
  \bibfield  {author} {\bibinfo {author} {\bibfnamefont {C.~S.}\ \bibnamefont {Dweck}},\ }\href {https://doi.org/10.4324/9781315783048} {\emph {\bibinfo {title} {{Self-theories}}}}\ (\bibinfo  {publisher} {Psychology Press},\ \bibinfo {year} {1999})\BibitemShut {NoStop}%
\bibitem [{\citenamefont {Hong}\ \emph {et~al.}(1999)\citenamefont {Hong}, \citenamefont {Chiu}, \citenamefont {Dweck}, \citenamefont {Lin},\ and\ \citenamefont {Wan}}]{Hong1999}%
  \BibitemOpen
  \bibfield  {author} {\bibinfo {author} {\bibfnamefont {Y.-y.}\ \bibnamefont {Hong}}, \bibinfo {author} {\bibfnamefont {C.-y.}\ \bibnamefont {Chiu}}, \bibinfo {author} {\bibfnamefont {C.~S.}\ \bibnamefont {Dweck}}, \bibinfo {author} {\bibfnamefont {D.~M.-S.}\ \bibnamefont {Lin}},\ and\ \bibinfo {author} {\bibfnamefont {W.}~\bibnamefont {Wan}},\ }\bibfield  {title} {\bibinfo {title} {{Implicit theories, attributions, and coping: A meaning system approach.}},\ }\href {https://doi.org/10.1037/0022-3514.77.3.588} {\bibfield  {journal} {\bibinfo  {journal} {Journal of Personality and Social Psychology}\ }\textbf {\bibinfo {volume} {77}},\ \bibinfo {pages} {588} (\bibinfo {year} {1999})}\BibitemShut {NoStop}%
\bibitem [{\citenamefont {Heine}\ \emph {et~al.}(2001)\citenamefont {Heine}, \citenamefont {Kitayama}, \citenamefont {Lehman}, \citenamefont {Takata}, \citenamefont {Ide}, \citenamefont {Leung},\ and\ \citenamefont {Matsumoto}}]{Heine2001}%
  \BibitemOpen
  \bibfield  {author} {\bibinfo {author} {\bibfnamefont {S.~J.}\ \bibnamefont {Heine}}, \bibinfo {author} {\bibfnamefont {S.}~\bibnamefont {Kitayama}}, \bibinfo {author} {\bibfnamefont {D.~R.}\ \bibnamefont {Lehman}}, \bibinfo {author} {\bibfnamefont {T.}~\bibnamefont {Takata}}, \bibinfo {author} {\bibfnamefont {E.}~\bibnamefont {Ide}}, \bibinfo {author} {\bibfnamefont {C.}~\bibnamefont {Leung}},\ and\ \bibinfo {author} {\bibfnamefont {H.}~\bibnamefont {Matsumoto}},\ }\bibfield  {title} {\bibinfo {title} {{Divergent consequences of success and failure in Japan and North America: An investigation of self-improving motivations and malleable selves.}},\ }\href {https://doi.org/10.1037/0022-3514.81.4.599} {\bibfield  {journal} {\bibinfo  {journal} {Journal of Personality and Social Psychology}\ }\textbf {\bibinfo {volume} {81}},\ \bibinfo {pages} {599} (\bibinfo {year} {2001})}\BibitemShut {NoStop}%
\bibitem [{\citenamefont {Blackwell}\ \emph {et~al.}(2007)\citenamefont {Blackwell}, \citenamefont {Trzesniewski},\ and\ \citenamefont {Dweck}}]{Blackwell2007}%
  \BibitemOpen
  \bibfield  {author} {\bibinfo {author} {\bibfnamefont {L.~S.}\ \bibnamefont {Blackwell}}, \bibinfo {author} {\bibfnamefont {K.~H.}\ \bibnamefont {Trzesniewski}},\ and\ \bibinfo {author} {\bibfnamefont {C.~S.}\ \bibnamefont {Dweck}},\ }\bibfield  {title} {\bibinfo {title} {{Implicit Theories of Intelligence Predict Achievement Across an Adolescent Transition: A Longitudinal Study and an Intervention}},\ }\href {https://doi.org/10.1111/j.1467-8624.2007.00995.x} {\bibfield  {journal} {\bibinfo  {journal} {Child Development}\ }\textbf {\bibinfo {volume} {78}},\ \bibinfo {pages} {246} (\bibinfo {year} {2007})}\BibitemShut {NoStop}%
\bibitem [{\citenamefont {Burnette}\ \emph {et~al.}(2013)\citenamefont {Burnette}, \citenamefont {O'Boyle}, \citenamefont {VanEpps}, \citenamefont {Pollack},\ and\ \citenamefont {Finkel}}]{Burnette2013}%
  \BibitemOpen
  \bibfield  {author} {\bibinfo {author} {\bibfnamefont {J.~L.}\ \bibnamefont {Burnette}}, \bibinfo {author} {\bibfnamefont {E.~H.}\ \bibnamefont {O'Boyle}}, \bibinfo {author} {\bibfnamefont {E.~M.}\ \bibnamefont {VanEpps}}, \bibinfo {author} {\bibfnamefont {J.~M.}\ \bibnamefont {Pollack}},\ and\ \bibinfo {author} {\bibfnamefont {E.~J.}\ \bibnamefont {Finkel}},\ }\bibfield  {title} {\bibinfo {title} {{Mind-sets matter: A meta-analytic review of implicit theories and self-regulation.}},\ }\href {https://doi.org/10.1037/a0029531} {\bibfield  {journal} {\bibinfo  {journal} {Psychological Bulletin}\ }\textbf {\bibinfo {volume} {139}},\ \bibinfo {pages} {655} (\bibinfo {year} {2013})}\BibitemShut {NoStop}%
\bibitem [{\citenamefont {Paunesku}\ \emph {et~al.}(2015)\citenamefont {Paunesku}, \citenamefont {Walton}, \citenamefont {Romero}, \citenamefont {Smith}, \citenamefont {Yeager},\ and\ \citenamefont {Dweck}}]{Paunesku2015}%
  \BibitemOpen
  \bibfield  {author} {\bibinfo {author} {\bibfnamefont {D.}~\bibnamefont {Paunesku}}, \bibinfo {author} {\bibfnamefont {G.~M.}\ \bibnamefont {Walton}}, \bibinfo {author} {\bibfnamefont {C.}~\bibnamefont {Romero}}, \bibinfo {author} {\bibfnamefont {E.~N.}\ \bibnamefont {Smith}}, \bibinfo {author} {\bibfnamefont {D.~S.}\ \bibnamefont {Yeager}},\ and\ \bibinfo {author} {\bibfnamefont {C.~S.}\ \bibnamefont {Dweck}},\ }\bibfield  {title} {\bibinfo {title} {{Mind-Set Interventions Are a Scalable Treatment for Academic Underachievement}},\ }\href {https://doi.org/10.1177/0956797615571017} {\bibfield  {journal} {\bibinfo  {journal} {Psychological Science}\ }\textbf {\bibinfo {volume} {26}},\ \bibinfo {pages} {784} (\bibinfo {year} {2015})}\BibitemShut {NoStop}%
\bibitem [{\citenamefont {Smiley}\ \emph {et~al.}(2016)\citenamefont {Smiley}, \citenamefont {Buttitta}, \citenamefont {Chung}, \citenamefont {Dubon},\ and\ \citenamefont {Chang}}]{Smiley2016}%
  \BibitemOpen
  \bibfield  {author} {\bibinfo {author} {\bibfnamefont {P.~A.}\ \bibnamefont {Smiley}}, \bibinfo {author} {\bibfnamefont {K.~V.}\ \bibnamefont {Buttitta}}, \bibinfo {author} {\bibfnamefont {S.~Y.}\ \bibnamefont {Chung}}, \bibinfo {author} {\bibfnamefont {V.~X.}\ \bibnamefont {Dubon}},\ and\ \bibinfo {author} {\bibfnamefont {L.~K.}\ \bibnamefont {Chang}},\ }\bibfield  {title} {\bibinfo {title} {{Mediation models of implicit theories and achievement goals predict planning and withdrawal after failure}},\ }\href {https://doi.org/10.1007/s11031-016-9575-5} {\bibfield  {journal} {\bibinfo  {journal} {Motivation and Emotion}\ }\textbf {\bibinfo {volume} {40}},\ \bibinfo {pages} {878} (\bibinfo {year} {2016})}\BibitemShut {NoStop}%
\bibitem [{\citenamefont {Bahn{\'{i}}k}\ and\ \citenamefont {Vranka}(2017)}]{Bahnik2017}%
  \BibitemOpen
  \bibfield  {author} {\bibinfo {author} {\bibfnamefont {{\v{S}}.}~\bibnamefont {Bahn{\'{i}}k}}\ and\ \bibinfo {author} {\bibfnamefont {M.~A.}\ \bibnamefont {Vranka}},\ }\bibfield  {title} {\bibinfo {title} {{Growth mindset is not associated with scholastic aptitude in a large sample of university applicants}},\ }\href {https://doi.org/10.1016/j.paid.2017.05.046} {\bibfield  {journal} {\bibinfo  {journal} {Personality and Individual Differences}\ }\textbf {\bibinfo {volume} {117}},\ \bibinfo {pages} {139} (\bibinfo {year} {2017})}\BibitemShut {NoStop}%
\bibitem [{\citenamefont {S{\'{a}}nchez}\ and\ \citenamefont {R{\'{i}}os}(2020)}]{Sanchez2020}%
  \BibitemOpen
  \bibfield  {author} {\bibinfo {author} {\bibfnamefont {A.~X.}\ \bibnamefont {S{\'{a}}nchez}}\ and\ \bibinfo {author} {\bibfnamefont {L.}~\bibnamefont {R{\'{i}}os}},\ }\bibfield  {title} {\bibinfo {title} {{Analysis of student perceptions of classroom structure, belongingness, and motivation in an introductory physics course}},\ }in\ \href {https://doi.org/10.1119/perc.2020.pr.Sanchez} {\emph {\bibinfo {booktitle} {2020 Physics Education Research Conference Proceedings}}}\ (\bibinfo  {publisher} {American Association of Physics Teachers},\ \bibinfo {year} {2020})\ pp.\ \bibinfo {pages} {454--459}\BibitemShut {NoStop}%
\bibitem [{\citenamefont {Hoyert}\ and\ \citenamefont {O'Dell}(2008)}]{Hoyert2008}%
  \BibitemOpen
  \bibfield  {author} {\bibinfo {author} {\bibfnamefont {M.}~\bibnamefont {Hoyert}}\ and\ \bibinfo {author} {\bibfnamefont {C.}~\bibnamefont {O'Dell}},\ }\bibfield  {title} {\bibinfo {title} {{Goal Orientation and the Aftermath of an Academic Failure}},\ }\href {https://doi.org/10.18848/1447-9494/CGP/v15i03/45663} {\bibfield  {journal} {\bibinfo  {journal} {The International Journal of Learning: Annual Review}\ }\textbf {\bibinfo {volume} {15}},\ \bibinfo {pages} {245} (\bibinfo {year} {2008})}\BibitemShut {NoStop}%
\bibitem [{\citenamefont {Lisberg}\ and\ \citenamefont {Woods}(2018)}]{Lisberg2018}%
  \BibitemOpen
  \bibfield  {author} {\bibinfo {author} {\bibfnamefont {A.}~\bibnamefont {Lisberg}}\ and\ \bibinfo {author} {\bibfnamefont {B.}~\bibnamefont {Woods}},\ }\bibfield  {title} {\bibinfo {title} {{Mentorship, Mindset and Learning Strategies: An Integrative Approach to Increasing Underrepresented Minority Student Retention in a STEM Undergraduate Program.}},\ }\href {https://www.jstem.org/jstem/index.php/JSTEM/
  article/view/2280} {\bibfield  {journal} {\bibinfo  {journal} {Journal of STEM Education: Innovations and Research}\ }\textbf {\bibinfo {volume} {19}},\ \bibinfo {pages} {14} (\bibinfo {year} {2018})}\BibitemShut {NoStop}%
\bibitem [{\citenamefont {Beatty}\ \emph {et~al.}(2020)\citenamefont {Beatty}, \citenamefont {Sedberry}, \citenamefont {Gerace}, \citenamefont {Strickhouser}, \citenamefont {Elobeid}, \citenamefont {Kane}, \citenamefont {Beatty}, \citenamefont {Sedberry}, \citenamefont {Gerace}, \citenamefont {Strickhouser}, \citenamefont {Elobeid},\ and\ \citenamefont {Kane}}]{Beatty2019}%
  \BibitemOpen
  \bibfield  {author} {\bibinfo {author} {\bibfnamefont {I.~D.}\ \bibnamefont {Beatty}}, \bibinfo {author} {\bibfnamefont {S.~J.}\ \bibnamefont {Sedberry}}, \bibinfo {author} {\bibfnamefont {W.~J.}\ \bibnamefont {Gerace}}, \bibinfo {author} {\bibfnamefont {J.~E.}\ \bibnamefont {Strickhouser}}, \bibinfo {author} {\bibfnamefont {M.~A.}\ \bibnamefont {Elobeid}}, \bibinfo {author} {\bibfnamefont {M.~J.}\ \bibnamefont {Kane}}, \bibinfo {author} {\bibfnamefont {I.~D.}\ \bibnamefont {Beatty}}, \bibinfo {author} {\bibfnamefont {S.~J.}\ \bibnamefont {Sedberry}}, \bibinfo {author} {\bibfnamefont {W.~J.}\ \bibnamefont {Gerace}}, \bibinfo {author} {\bibfnamefont {J.~E.}\ \bibnamefont {Strickhouser}}, \bibinfo {author} {\bibfnamefont {M.~A.}\ \bibnamefont {Elobeid}},\ and\ \bibinfo {author} {\bibfnamefont {M.~J.}\ \bibnamefont {Kane}},\ }\bibfield  {title} {\bibinfo {title} {{Improving STEM self-efficacy with a scalable classroom intervention targeting growth mindset and success attribution}},\ }in\ \href
  {https://doi.org/10.1119/perc.2019.pr.Beatty} {\emph {\bibinfo {booktitle} {2019 Physics Education Research Conference Proceedings}}}\ (\bibinfo  {publisher} {American Association of Physics Teachers},\ \bibinfo {year} {2020})\ pp.\ \bibinfo {pages} {44--50}\BibitemShut {NoStop}%
\bibitem [{\citenamefont {Nallapothula}\ \emph {et~al.}(2020)\citenamefont {Nallapothula}, \citenamefont {Lozano}, \citenamefont {Han}, \citenamefont {Herrera}, \citenamefont {Sayson}, \citenamefont {Levis-Fitzgerald},\ and\ \citenamefont {Maloy}}]{Nallapothula2020}%
  \BibitemOpen
  \bibfield  {author} {\bibinfo {author} {\bibfnamefont {D.}~\bibnamefont {Nallapothula}}, \bibinfo {author} {\bibfnamefont {J.~B.}\ \bibnamefont {Lozano}}, \bibinfo {author} {\bibfnamefont {S.}~\bibnamefont {Han}}, \bibinfo {author} {\bibfnamefont {C.}~\bibnamefont {Herrera}}, \bibinfo {author} {\bibfnamefont {H.~W.}\ \bibnamefont {Sayson}}, \bibinfo {author} {\bibfnamefont {M.}~\bibnamefont {Levis-Fitzgerald}},\ and\ \bibinfo {author} {\bibfnamefont {J.}~\bibnamefont {Maloy}},\ }\bibfield  {title} {\bibinfo {title} {{M-LoCUS: A Scalable Intervention Enhances Growth Mindset and Internal Locus of Control in Undergraduate Students in STEM}},\ }\href {https://doi.org/10.1128/jmbe.v21i2.1987} \bibfield  {journal} {\bibinfo  {journal} {Journal of Microbiology {\&} Biology Education}\ }\textbf {\bibinfo {volume} {21}},\ \bibinfo {pages} {30}  (\bibinfo {year} {2020})\BibitemShut {NoStop}%
\bibitem [{\citenamefont {Chambers}\ \emph {et~al.}(2022)\citenamefont {Chambers}, \citenamefont {Lowe},\ and\ \citenamefont {Muldrow}}]{Chambers2022}%
  \BibitemOpen
  \bibfield  {author} {\bibinfo {author} {\bibfnamefont {B.}~\bibnamefont {Chambers}}, \bibinfo {author} {\bibfnamefont {J.}~\bibnamefont {Lowe}},\ and\ \bibinfo {author} {\bibfnamefont {L.}~\bibnamefont {Muldrow}},\ }\bibfield  {title} {\bibinfo {title} {{Dissemination of Growth Mindset Principles and Attitudes in the Division of Science and Mathematics at a Liberal Arts College.}},\ }\href {https://www.jstem.org/jstem/index.php/JSTEM/article/view/2536} {\bibfield  {journal} {\bibinfo  {journal} {Journal of STEM Education: Innovations and Research}\ }\textbf {\bibinfo {volume} {23}},\ \bibinfo {pages} {35} (\bibinfo {year} {2022})}\BibitemShut {NoStop}%
\bibitem [{\citenamefont {Aronson}\ \emph {et~al.}(2002)\citenamefont {Aronson}, \citenamefont {Fried},\ and\ \citenamefont {Good}}]{Aronson2002}%
  \BibitemOpen
  \bibfield  {author} {\bibinfo {author} {\bibfnamefont {J.}~\bibnamefont {Aronson}}, \bibinfo {author} {\bibfnamefont {C.~B.}\ \bibnamefont {Fried}},\ and\ \bibinfo {author} {\bibfnamefont {C.}~\bibnamefont {Good}},\ }\bibfield  {title} {\bibinfo {title} {{Reducing the Effects of Stereotype Threat on African American College Students by Shaping Theories of Intelligence}},\ }\href {https://doi.org/10.1006/jesp.2001.1491} {\bibfield  {journal} {\bibinfo  {journal} {Journal of Experimental Social Psychology}\ }\textbf {\bibinfo {volume} {38}},\ \bibinfo {pages} {113} (\bibinfo {year} {2002})}\BibitemShut {NoStop}%
\bibitem [{\citenamefont {Good}\ \emph {et~al.}(2003)\citenamefont {Good}, \citenamefont {Aronson},\ and\ \citenamefont {Inzlicht}}]{Good2003}%
  \BibitemOpen
  \bibfield  {author} {\bibinfo {author} {\bibfnamefont {C.}~\bibnamefont {Good}}, \bibinfo {author} {\bibfnamefont {J.}~\bibnamefont {Aronson}},\ and\ \bibinfo {author} {\bibfnamefont {M.}~\bibnamefont {Inzlicht}},\ }\bibfield  {title} {\bibinfo {title} {{Improving adolescents' standardized test performance: An intervention to reduce the effects of stereotype threat}},\ }\href {https://doi.org/10.1016/j.appdev.2003.09.002} {\bibfield  {journal} {\bibinfo  {journal} {Journal of Applied Developmental Psychology}\ }\textbf {\bibinfo {volume} {24}},\ \bibinfo {pages} {645} (\bibinfo {year} {2003})}\BibitemShut {NoStop}%
\bibitem [{\citenamefont {Yeager}\ and\ \citenamefont {Walton}(2011)}]{Yeager2011}%
  \BibitemOpen
  \bibfield  {author} {\bibinfo {author} {\bibfnamefont {D.~S.}\ \bibnamefont {Yeager}}\ and\ \bibinfo {author} {\bibfnamefont {G.~M.}\ \bibnamefont {Walton}},\ }\bibfield  {title} {\bibinfo {title} {{Social-Psychological Interventions in Education: They're Not Magic}},\ }\href {http://www.jstor.org/stable/23014370} {\bibfield  {journal} {\bibinfo  {journal} {Review of Educational Research}\ }\textbf {\bibinfo {volume} {81}},\ \bibinfo {pages} {267} (\bibinfo {year} {2011})}\BibitemShut {NoStop}%
\bibitem [{\citenamefont {Yeager}\ \emph {et~al.}(2019)\citenamefont {Yeager}, \citenamefont {Hanselman}, \citenamefont {Walton}, \citenamefont {Murray}, \citenamefont {Crosnoe}, \citenamefont {Muller}, \citenamefont {Tipton}, \citenamefont {Schneider}, \citenamefont {Hulleman}, \citenamefont {Hinojosa}, \citenamefont {Paunesku}, \citenamefont {Romero}, \citenamefont {Flint}, \citenamefont {Roberts}, \citenamefont {Trott}, \citenamefont {Iachan}, \citenamefont {Buontempo}, \citenamefont {Yang}, \citenamefont {Carvalho}, \citenamefont {Hahn}, \citenamefont {Gopalan}, \citenamefont {Mhatre}, \citenamefont {Ferguson}, \citenamefont {Duckworth},\ and\ \citenamefont {Dweck}}]{Yeager2019}%
  \BibitemOpen
  \bibfield  {author} {\bibinfo {author} {\bibfnamefont {D.~S.}\ \bibnamefont {Yeager}}, \bibinfo {author} {\bibfnamefont {P.}~\bibnamefont {Hanselman}}, \bibinfo {author} {\bibfnamefont {G.~M.}\ \bibnamefont {Walton}}, \bibinfo {author} {\bibfnamefont {J.~S.}\ \bibnamefont {Murray}}, \bibinfo {author} {\bibfnamefont {R.}~\bibnamefont {Crosnoe}}, \bibinfo {author} {\bibfnamefont {C.}~\bibnamefont {Muller}}, \bibinfo {author} {\bibfnamefont {E.}~\bibnamefont {Tipton}}, \bibinfo {author} {\bibfnamefont {B.}~\bibnamefont {Schneider}}, \bibinfo {author} {\bibfnamefont {C.~S.}\ \bibnamefont {Hulleman}}, \bibinfo {author} {\bibfnamefont {C.~P.}\ \bibnamefont {Hinojosa}}, \bibinfo {author} {\bibfnamefont {D.}~\bibnamefont {Paunesku}}, \bibinfo {author} {\bibfnamefont {C.}~\bibnamefont {Romero}}, \bibinfo {author} {\bibfnamefont {K.}~\bibnamefont {Flint}}, \bibinfo {author} {\bibfnamefont {A.}~\bibnamefont {Roberts}}, \bibinfo {author} {\bibfnamefont {J.}~\bibnamefont {Trott}}, \bibinfo {author} {\bibfnamefont
  {R.}~\bibnamefont {Iachan}}, \bibinfo {author} {\bibfnamefont {J.}~\bibnamefont {Buontempo}}, \bibinfo {author} {\bibfnamefont {S.~M.}\ \bibnamefont {Yang}}, \bibinfo {author} {\bibfnamefont {C.~M.}\ \bibnamefont {Carvalho}}, \bibinfo {author} {\bibfnamefont {P.~R.}\ \bibnamefont {Hahn}}, \bibinfo {author} {\bibfnamefont {M.}~\bibnamefont {Gopalan}}, \bibinfo {author} {\bibfnamefont {P.}~\bibnamefont {Mhatre}}, \bibinfo {author} {\bibfnamefont {R.}~\bibnamefont {Ferguson}}, \bibinfo {author} {\bibfnamefont {A.~L.}\ \bibnamefont {Duckworth}},\ and\ \bibinfo {author} {\bibfnamefont {C.~S.}\ \bibnamefont {Dweck}},\ }\bibfield  {title} {\bibinfo {title} {{A national experiment reveals where a growth mindset improves achievement}},\ }\href {https://doi.org/10.1038/s41586-019-1466-y} {\bibfield  {journal} {\bibinfo  {journal} {Nature}\ }\textbf {\bibinfo {volume} {573}},\ \bibinfo {pages} {364} (\bibinfo {year} {2019})}\BibitemShut {NoStop}%
\bibitem [{\citenamefont {Orosz}\ \emph {et~al.}(2017)\citenamefont {Orosz}, \citenamefont {P{\'{e}}ter-Szarka}, \citenamefont {Bőthe}, \citenamefont {T{\'{o}}th-Kir{\'{a}}ly},\ and\ \citenamefont {Berger}}]{Orosz2017}%
  \BibitemOpen
  \bibfield  {author} {\bibinfo {author} {\bibfnamefont {G.}~\bibnamefont {Orosz}}, \bibinfo {author} {\bibfnamefont {S.}~\bibnamefont {P{\'{e}}ter-Szarka}}, \bibinfo {author} {\bibfnamefont {B.}~\bibnamefont {Bőthe}}, \bibinfo {author} {\bibfnamefont {I.}~\bibnamefont {T{\'{o}}th-Kir{\'{a}}ly}},\ and\ \bibinfo {author} {\bibfnamefont {R.}~\bibnamefont {Berger}},\ }\bibfield  {title} {\bibinfo {title} {{How Not to Do a Mindset Intervention: Learning from a Mindset Intervention among Students with Good Grades}},\ }\href {https://doi.org/10.3389/fpsyg.2017.00311} {\bibfield  {journal} {\bibinfo  {journal} {Frontiers in Psychology}\ }\textbf {\bibinfo {volume} {8}},\ \bibinfo {pages} {311} (\bibinfo {year} {2017})}\BibitemShut {NoStop}%
\bibitem [{\citenamefont {Sisk}\ \emph {et~al.}(2018)\citenamefont {Sisk}, \citenamefont {Burgoyne}, \citenamefont {Sun}, \citenamefont {Butler},\ and\ \citenamefont {Macnamara}}]{Sisk2018}%
  \BibitemOpen
  \bibfield  {author} {\bibinfo {author} {\bibfnamefont {V.~F.}\ \bibnamefont {Sisk}}, \bibinfo {author} {\bibfnamefont {A.~P.}\ \bibnamefont {Burgoyne}}, \bibinfo {author} {\bibfnamefont {J.}~\bibnamefont {Sun}}, \bibinfo {author} {\bibfnamefont {J.~L.}\ \bibnamefont {Butler}},\ and\ \bibinfo {author} {\bibfnamefont {B.~N.}\ \bibnamefont {Macnamara}},\ }\bibfield  {title} {\bibinfo {title} {{To What Extent and Under Which Circumstances Are Growth Mind-Sets Important to Academic Achievement? Two Meta-Analyses}},\ }\href {https://doi.org/10.1177/0956797617739704} {\bibfield  {journal} {\bibinfo  {journal} {Psychological Science}\ }\textbf {\bibinfo {volume} {29}},\ \bibinfo {pages} {549} (\bibinfo {year} {2018})}\BibitemShut {NoStop}%
\bibitem [{\citenamefont {Wang}\ \emph {et~al.}(2021)\citenamefont {Wang}, \citenamefont {Rocabado}, \citenamefont {Lewis},\ and\ \citenamefont {Lewis}}]{Wang2021}%
  \BibitemOpen
  \bibfield  {author} {\bibinfo {author} {\bibfnamefont {Y.}~\bibnamefont {Wang}}, \bibinfo {author} {\bibfnamefont {G.~A.}\ \bibnamefont {Rocabado}}, \bibinfo {author} {\bibfnamefont {J.~E.}\ \bibnamefont {Lewis}},\ and\ \bibinfo {author} {\bibfnamefont {S.~E.}\ \bibnamefont {Lewis}},\ }\bibfield  {title} {\bibinfo {title} {{Prompts to Promote Success: Evaluating Utility Value and Growth Mindset Interventions on General Chemistry Students' Attitude and Academic Performance}},\ }\href {https://doi.org/10.1021/acs.jchemed.0c01497} {\bibfield  {journal} {\bibinfo  {journal} {Journal of Chemical Education}\ }\textbf {\bibinfo {volume} {98}},\ \bibinfo {pages} {1476} (\bibinfo {year} {2021})}\BibitemShut {NoStop}%
\bibitem [{\citenamefont {Fink}\ \emph {et~al.}(2018)\citenamefont {Fink}, \citenamefont {Cahill}, \citenamefont {McDaniel}, \citenamefont {Hoffman},\ and\ \citenamefont {Frey}}]{Fink2018}%
  \BibitemOpen
  \bibfield  {author} {\bibinfo {author} {\bibfnamefont {A.}~\bibnamefont {Fink}}, \bibinfo {author} {\bibfnamefont {M.~J.}\ \bibnamefont {Cahill}}, \bibinfo {author} {\bibfnamefont {M.~A.}\ \bibnamefont {McDaniel}}, \bibinfo {author} {\bibfnamefont {A.}~\bibnamefont {Hoffman}},\ and\ \bibinfo {author} {\bibfnamefont {R.~F.}\ \bibnamefont {Frey}},\ }\bibfield  {title} {\bibinfo {title} {{Improving general chemistry performance through a growth mindset intervention: selective effects on underrepresented minorities}},\ }\href {https://doi.org/10.1039/C7RP00244K} {\bibfield  {journal} {\bibinfo  {journal} {Chemistry Education Research and Practice}\ }\textbf {\bibinfo {volume} {19}},\ \bibinfo {pages} {783} (\bibinfo {year} {2018})}\BibitemShut {NoStop}%
\bibitem [{\citenamefont {Burnette}\ \emph {et~al.}(2022)\citenamefont {Burnette}, \citenamefont {Billingsley}, \citenamefont {Banks}, \citenamefont {Knouse}, \citenamefont {Hoyt}, \citenamefont {Pollack},\ and\ \citenamefont {Simon}}]{burnette2022}%
  \BibitemOpen
  \bibfield  {author} {\bibinfo {author} {\bibfnamefont {J.~L.}\ \bibnamefont {Burnette}}, \bibinfo {author} {\bibfnamefont {J.}~\bibnamefont {Billingsley}}, \bibinfo {author} {\bibfnamefont {G.~C.}\ \bibnamefont {Banks}}, \bibinfo {author} {\bibfnamefont {L.~E.}\ \bibnamefont {Knouse}}, \bibinfo {author} {\bibfnamefont {C.~L.}\ \bibnamefont {Hoyt}}, \bibinfo {author} {\bibfnamefont {J.~M.}\ \bibnamefont {Pollack}},\ and\ \bibinfo {author} {\bibfnamefont {S.}~\bibnamefont {Simon}},\ }\bibfield  {title} {\bibinfo {title} {{A systematic review and meta-analysis of growth mindset interventions: For whom, how, and why might such interventions work?}},\ }\bibfield  {journal} {\bibinfo  {journal} {Psychological Bulletin}\ }\href {https://doi.org/10.1037/bul0000368} {10.1037/bul0000368} (\bibinfo {year} {2022})\BibitemShut {NoStop}%
\bibitem [{\citenamefont {Kinlaw}\ and\ \citenamefont {Kurtz-Costes}(2003)}]{Kinlaw2003}%
  \BibitemOpen
  \bibfield  {author} {\bibinfo {author} {\bibfnamefont {C.~R.}\ \bibnamefont {Kinlaw}}\ and\ \bibinfo {author} {\bibfnamefont {B.}~\bibnamefont {Kurtz-Costes}},\ }\bibfield  {title} {\bibinfo {title} {{The development of children's beliefs about intelligence}},\ }\href {https://doi.org/10.1016/S0273-2297(03)00010-8} {\bibfield  {journal} {\bibinfo  {journal} {Developmental Review}\ }\textbf {\bibinfo {volume} {23}},\ \bibinfo {pages} {125} (\bibinfo {year} {2003})}\BibitemShut {NoStop}%
\bibitem [{\citenamefont {Gonida}\ \emph {et~al.}(2006)\citenamefont {Gonida}, \citenamefont {Kiosseoglou},\ and\ \citenamefont {Leondari}}]{Gonida2006}%
  \BibitemOpen
  \bibfield  {author} {\bibinfo {author} {\bibfnamefont {E.}~\bibnamefont {Gonida}}, \bibinfo {author} {\bibfnamefont {G.}~\bibnamefont {Kiosseoglou}},\ and\ \bibinfo {author} {\bibfnamefont {A.}~\bibnamefont {Leondari}},\ }\bibfield  {title} {\bibinfo {title} {{Implicit Theories of Intelligence, Perceived Academic Competence, and School Achievement: Testing Alternative Models}},\ }\href {https://doi.org/10.2307/20445336} {\bibfield  {journal} {\bibinfo  {journal} {The American Journal of Psychology}\ }\textbf {\bibinfo {volume} {119}},\ \bibinfo {pages} {223} (\bibinfo {year} {2006})}\BibitemShut {NoStop}%
\bibitem [{\citenamefont {Gunderson}\ \emph {et~al.}(2017)\citenamefont {Gunderson}, \citenamefont {Hamdan}, \citenamefont {Sorhagen},\ and\ \citenamefont {D'Esterre}}]{Gunderson2017}%
  \BibitemOpen
  \bibfield  {author} {\bibinfo {author} {\bibfnamefont {E.~A.}\ \bibnamefont {Gunderson}}, \bibinfo {author} {\bibfnamefont {N.}~\bibnamefont {Hamdan}}, \bibinfo {author} {\bibfnamefont {N.~S.}\ \bibnamefont {Sorhagen}},\ and\ \bibinfo {author} {\bibfnamefont {A.~P.}\ \bibnamefont {D'Esterre}},\ }\bibfield  {title} {\bibinfo {title} {{Who needs innate ability to succeed in math and literacy? Academic-domain-specific theories of intelligence about peers versus adults.}},\ }\href {https://doi.org/10.1037/dev0000282} {\bibfield  {journal} {\bibinfo  {journal} {Developmental Psychology}\ }\textbf {\bibinfo {volume} {53}},\ \bibinfo {pages} {1188} (\bibinfo {year} {2017})}\BibitemShut {NoStop}%
\bibitem [{\citenamefont {Robins}\ and\ \citenamefont {Pals}(2002)}]{Robins2002}%
  \BibitemOpen
  \bibfield  {author} {\bibinfo {author} {\bibfnamefont {R.~W.}\ \bibnamefont {Robins}}\ and\ \bibinfo {author} {\bibfnamefont {J.~L.}\ \bibnamefont {Pals}},\ }\bibfield  {title} {\bibinfo {title} {{Implicit Self-Theories in the Academic Domain: Implications for Goal Orientation, Attributions, Affect, and Self-Esteem Change}},\ }\href {https://doi.org/10.1080/15298860290106805} {\bibfield  {journal} {\bibinfo  {journal} {Self and Identity}\ }\textbf {\bibinfo {volume} {1}},\ \bibinfo {pages} {313} (\bibinfo {year} {2002})}\BibitemShut {NoStop}%
\bibitem [{\citenamefont {Dai}\ and\ \citenamefont {Cromley}(2014)}]{Dai2014}%
  \BibitemOpen
  \bibfield  {author} {\bibinfo {author} {\bibfnamefont {T.}~\bibnamefont {Dai}}\ and\ \bibinfo {author} {\bibfnamefont {J.~G.}\ \bibnamefont {Cromley}},\ }\bibfield  {title} {\bibinfo {title} {{Changes in implicit theories of ability in biology and dropout from STEM majors: A latent growth curve approach}},\ }\href {https://doi.org/10.1016/j.cedpsych.2014.06.003} {\bibfield  {journal} {\bibinfo  {journal} {Contemporary Educational Psychology}\ }\textbf {\bibinfo {volume} {39}},\ \bibinfo {pages} {233} (\bibinfo {year} {2014})}\BibitemShut {NoStop}%
\bibitem [{\citenamefont {Scott}\ and\ \citenamefont {Ghinea}(2014)}]{Scott2014}%
  \BibitemOpen
  \bibfield  {author} {\bibinfo {author} {\bibfnamefont {M.~J.}\ \bibnamefont {Scott}}\ and\ \bibinfo {author} {\bibfnamefont {G.}~\bibnamefont {Ghinea}},\ }\bibfield  {title} {\bibinfo {title} {{On the Domain-Specificity of Mindsets: The Relationship Between Aptitude Beliefs and Programming Practice}},\ }\href {https://doi.org/10.1109/TE.2013.2288700} {\bibfield  {journal} {\bibinfo  {journal} {IEEE Transactions on Education}\ }\textbf {\bibinfo {volume} {57}},\ \bibinfo {pages} {169} (\bibinfo {year} {2014})}\BibitemShut {NoStop}%
\bibitem [{\citenamefont {Flanigan}\ \emph {et~al.}(2017)\citenamefont {Flanigan}, \citenamefont {Peteranetz}, \citenamefont {Shell},\ and\ \citenamefont {Soh}}]{Flanigan2017}%
  \BibitemOpen
  \bibfield  {author} {\bibinfo {author} {\bibfnamefont {A.~E.}\ \bibnamefont {Flanigan}}, \bibinfo {author} {\bibfnamefont {M.~S.}\ \bibnamefont {Peteranetz}}, \bibinfo {author} {\bibfnamefont {D.~F.}\ \bibnamefont {Shell}},\ and\ \bibinfo {author} {\bibfnamefont {L.-K.}\ \bibnamefont {Soh}},\ }\bibfield  {title} {\bibinfo {title} {{Implicit intelligence beliefs of computer science students: Exploring change across the semester}},\ }\href {https://doi.org/10.1016/j.cedpsych.2016.10.003} {\bibfield  {journal} {\bibinfo  {journal} {Contemporary Educational Psychology}\ }\textbf {\bibinfo {volume} {48}},\ \bibinfo {pages} {179} (\bibinfo {year} {2017})}\BibitemShut {NoStop}%
\bibitem [{\citenamefont {Shively}\ and\ \citenamefont {Ryan}(2013)}]{Shively2013}%
  \BibitemOpen
  \bibfield  {author} {\bibinfo {author} {\bibfnamefont {R.~L.}\ \bibnamefont {Shively}}\ and\ \bibinfo {author} {\bibfnamefont {C.~S.}\ \bibnamefont {Ryan}},\ }\bibfield  {title} {\bibinfo {title} {{Longitudinal changes in college math students' implicit theories of intelligence}},\ }\href {https://doi.org/10.1007/s11218-012-9208-0} {\bibfield  {journal} {\bibinfo  {journal} {Social Psychology of Education}\ }\textbf {\bibinfo {volume} {16}},\ \bibinfo {pages} {241} (\bibinfo {year} {2013})}\BibitemShut {NoStop}%
\bibitem [{\citenamefont {Malespina}\ \emph {et~al.}(2022)\citenamefont {Malespina}, \citenamefont {Schunn},\ and\ \citenamefont {Singh}}]{Malespina2022}%
  \BibitemOpen
  \bibfield  {author} {\bibinfo {author} {\bibfnamefont {A.}~\bibnamefont {Malespina}}, \bibinfo {author} {\bibfnamefont {C.~D.}\ \bibnamefont {Schunn}},\ and\ \bibinfo {author} {\bibfnamefont {C.}~\bibnamefont {Singh}},\ }\bibfield  {title} {\bibinfo {title} {{To whom do students believe a growth mindset applies?}},\ }in\ \href {https://doi.org/10.1119/perc.2022.pr.Malespina} {\emph {\bibinfo {booktitle} {2022 Physics Education Research Conference Proceedings}}}\ (\bibinfo  {publisher} {American Association of Physics Teachers},\ \bibinfo {year} {2022})\ pp.\ \bibinfo {pages} {292--297}\BibitemShut {NoStop}%
\bibitem [{\citenamefont {Malespina}\ \emph {et~al.}(2023)\citenamefont {Malespina}, \citenamefont {Schunn},\ and\ \citenamefont {Singh}}]{Malespina2023}%
  \BibitemOpen
  \bibfield  {author} {\bibinfo {author} {\bibfnamefont {A.}~\bibnamefont {Malespina}}, \bibinfo {author} {\bibfnamefont {C.~D.}\ \bibnamefont {Schunn}},\ and\ \bibinfo {author} {\bibfnamefont {C.}~\bibnamefont {Singh}},\ }\bibfield  {title} {\bibinfo {title} {{Bioscience students' internalized mindsets predict grades and reveal gender inequities in physics courses}},\ }\href {https://doi.org/10.1103/PhysRevPhysEducRes.19.020135} {\bibfield  {journal} {\bibinfo  {journal} {Physical Review Physics Education Research}\ }\textbf {\bibinfo {volume} {19}},\ \bibinfo {pages} {020135} (\bibinfo {year} {2023})}\BibitemShut {NoStop}%
\bibitem [{\citenamefont {Limeri}\ \emph {et~al.}(2020)\citenamefont {Limeri}, \citenamefont {Carter}, \citenamefont {Choe}, \citenamefont {Harper}, \citenamefont {Martin}, \citenamefont {Benton},\ and\ \citenamefont {Dolan}}]{Limeri2020}%
  \BibitemOpen
  \bibfield  {author} {\bibinfo {author} {\bibfnamefont {L.~B.}\ \bibnamefont {Limeri}}, \bibinfo {author} {\bibfnamefont {N.~T.}\ \bibnamefont {Carter}}, \bibinfo {author} {\bibfnamefont {J.}~\bibnamefont {Choe}}, \bibinfo {author} {\bibfnamefont {H.~G.}\ \bibnamefont {Harper}}, \bibinfo {author} {\bibfnamefont {H.~R.}\ \bibnamefont {Martin}}, \bibinfo {author} {\bibfnamefont {A.}~\bibnamefont {Benton}},\ and\ \bibinfo {author} {\bibfnamefont {E.~L.}\ \bibnamefont {Dolan}},\ }\bibfield  {title} {\bibinfo {title} {{Growing a growth mindset: characterizing how and why undergraduate students' mindsets change}},\ }\href {https://doi.org/10.1186/s40594-020-00227-2} {\bibfield  {journal} {\bibinfo  {journal} {International Journal of STEM Education}\ }\textbf {\bibinfo {volume} {7}},\ \bibinfo {pages} {35} (\bibinfo {year} {2020})}\BibitemShut {NoStop}%
\bibitem [{Note1()}]{Note1}%
  \BibitemOpen
  \bibinfo {note} {“Survey of Attitudes Toward Astronomy” by Michael Zeilik; http://www.flaguide.org/tools/attitude/astpr.htm}\BibitemShut {NoStop}%
\bibitem [{\citenamefont {Schau}\ \emph {et~al.}(1995)\citenamefont {Schau}, \citenamefont {Stevens}, \citenamefont {Dauphinee},\ and\ \citenamefont {Vecchio}}]{Schau1995}%
  \BibitemOpen
  \bibfield  {author} {\bibinfo {author} {\bibfnamefont {C.}~\bibnamefont {Schau}}, \bibinfo {author} {\bibfnamefont {J.}~\bibnamefont {Stevens}}, \bibinfo {author} {\bibfnamefont {T.~L.}\ \bibnamefont {Dauphinee}},\ and\ \bibinfo {author} {\bibfnamefont {A.~D.}\ \bibnamefont {Vecchio}},\ }\bibfield  {title} {\bibinfo {title} {{The Development and Validation of the Survey of Antitudes toward Statistics}},\ }\href {https://doi.org/10.1177/0013164495055005022} {\bibfield  {journal} {\bibinfo  {journal} {Educational and Psychological Measurement}\ }\textbf {\bibinfo {volume} {55}},\ \bibinfo {pages} {868} (\bibinfo {year} {1995})}\BibitemShut {NoStop}%
\bibitem [{\citenamefont {Zeilik}\ \emph {et~al.}(1997)\citenamefont {Zeilik}, \citenamefont {Schau}, \citenamefont {Mattern}, \citenamefont {Hall}, \citenamefont {Teague},\ and\ \citenamefont {Bisard}}]{Zeilik1997}%
  \BibitemOpen
  \bibfield  {author} {\bibinfo {author} {\bibfnamefont {M.}~\bibnamefont {Zeilik}}, \bibinfo {author} {\bibfnamefont {C.}~\bibnamefont {Schau}}, \bibinfo {author} {\bibfnamefont {N.}~\bibnamefont {Mattern}}, \bibinfo {author} {\bibfnamefont {S.}~\bibnamefont {Hall}}, \bibinfo {author} {\bibfnamefont {K.~W.}\ \bibnamefont {Teague}},\ and\ \bibinfo {author} {\bibfnamefont {W.}~\bibnamefont {Bisard}},\ }\bibfield  {title} {\bibinfo {title} {{Conceptual astronomy: A novel model for teaching postsecondary science courses}},\ }\href {https://doi.org/10.1119/1.18702} {\bibfield  {journal} {\bibinfo  {journal} {American Journal of Physics}\ }\textbf {\bibinfo {volume} {65}},\ \bibinfo {pages} {987} (\bibinfo {year} {1997})}\BibitemShut {NoStop}%
\bibitem [{\citenamefont {Zeilik}\ \emph {et~al.}(1999)\citenamefont {Zeilik}, \citenamefont {Schau},\ and\ \citenamefont {Mattern}}]{Zeilik1999}%
  \BibitemOpen
  \bibfield  {author} {\bibinfo {author} {\bibfnamefont {M.}~\bibnamefont {Zeilik}}, \bibinfo {author} {\bibfnamefont {C.}~\bibnamefont {Schau}},\ and\ \bibinfo {author} {\bibfnamefont {N.}~\bibnamefont {Mattern}},\ }\bibfield  {title} {\bibinfo {title} {{Conceptual astronomy. II. Replicating conceptual gains, probing attitude changes across three semesters}},\ }\href {https://doi.org/10.1119/1.19151} {\bibfield  {journal} {\bibinfo  {journal} {American Journal of Physics}\ }\textbf {\bibinfo {volume} {67}},\ \bibinfo {pages} {923} (\bibinfo {year} {1999})}\BibitemShut {NoStop}%
\bibitem [{Note2()}]{Note2}%
  \BibitemOpen
  \bibinfo {note} {“Mindset Quiz” by Emily Diehl, Educator and Author; https://classroom20.com/forum/topics/motivating-students-with?commentId=649749\%3AComment\%3A1167061}\BibitemShut {NoStop}%
\bibitem [{\citenamefont {Hacisalihoglu}\ \emph {et~al.}(2020)\citenamefont {Hacisalihoglu}, \citenamefont {Stephens}, \citenamefont {Stephens}, \citenamefont {Johnson},\ and\ \citenamefont {Edington}}]{Hacisalihoglu2020}%
  \BibitemOpen
  \bibfield  {author} {\bibinfo {author} {\bibfnamefont {G.}~\bibnamefont {Hacisalihoglu}}, \bibinfo {author} {\bibfnamefont {D.}~\bibnamefont {Stephens}}, \bibinfo {author} {\bibfnamefont {S.}~\bibnamefont {Stephens}}, \bibinfo {author} {\bibfnamefont {L.}~\bibnamefont {Johnson}},\ and\ \bibinfo {author} {\bibfnamefont {M.}~\bibnamefont {Edington}},\ }\bibfield  {title} {\bibinfo {title} {{Enhancing Undergraduate Student Success in STEM Fields through Growth-Mindset and Grit.}},\ }\href {https://eric.ed.gov/?id=EJ1272717} {\bibfield  {journal} {\bibinfo  {journal} {Education Sciences}\ }\textbf {\bibinfo {volume} {10}} (\bibinfo {year} {2020})}\BibitemShut {NoStop}%
\bibitem [{\citenamefont {Cook}\ \emph {et~al.}(2017)\citenamefont {Cook}, \citenamefont {Castillo}, \citenamefont {Gas},\ and\ \citenamefont {Artino}}]{Cook2017}%
  \BibitemOpen
  \bibfield  {author} {\bibinfo {author} {\bibfnamefont {D.~A.}\ \bibnamefont {Cook}}, \bibinfo {author} {\bibfnamefont {R.~M.}\ \bibnamefont {Castillo}}, \bibinfo {author} {\bibfnamefont {B.}~\bibnamefont {Gas}},\ and\ \bibinfo {author} {\bibfnamefont {A.~R.}\ \bibnamefont {Artino}},\ }\bibfield  {title} {\bibinfo {title} {{Measuring achievement goal motivation, mindsets and cognitive load: validation of three instruments' scores}},\ }\href {https://doi.org/10.1111/medu.13405} {\bibfield  {journal} {\bibinfo  {journal} {Medical Education}\ }\textbf {\bibinfo {volume} {51}},\ \bibinfo {pages} {1061} (\bibinfo {year} {2017})}\BibitemShut {NoStop}%
\bibitem [{\citenamefont {Troche}\ and\ \citenamefont {Kunz}(2020)}]{Troche2020}%
  \BibitemOpen
  \bibfield  {author} {\bibinfo {author} {\bibfnamefont {S.~J.}\ \bibnamefont {Troche}}\ and\ \bibinfo {author} {\bibfnamefont {A.}~\bibnamefont {Kunz}},\ }\bibfield  {title} {\bibinfo {title} {{The factorial structure and construct validity of a German translation of Dweck's Implicit Theories of Intelligence Scale under consideration of the wording effect.}},\ }\href {https://boris.unibe.ch/153429/} {\bibfield  {journal} {\bibinfo  {journal} {Psychological Test and Assessment Modeling}\ }\textbf {\bibinfo {volume} {62}},\ \bibinfo {pages} {386} (\bibinfo {year} {2020})}\BibitemShut {NoStop}%
\bibitem [{Note3()}]{Note3}%
  \BibitemOpen
  \bibinfo {note} {https://www.qualtrics.com}\BibitemShut {NoStop}%
\bibitem [{\citenamefont {Leslie}\ \emph {et~al.}(2015)\citenamefont {Leslie}, \citenamefont {Cimpian}, \citenamefont {Meyer},\ and\ \citenamefont {Freeland}}]{Leslie2015}%
  \BibitemOpen
  \bibfield  {author} {\bibinfo {author} {\bibfnamefont {S.-J.}\ \bibnamefont {Leslie}}, \bibinfo {author} {\bibfnamefont {A.}~\bibnamefont {Cimpian}}, \bibinfo {author} {\bibfnamefont {M.}~\bibnamefont {Meyer}},\ and\ \bibinfo {author} {\bibfnamefont {E.}~\bibnamefont {Freeland}},\ }\bibfield  {title} {\bibinfo {title} {{Expectations of brilliance underlie gender distributions across academic disciplines}},\ }\href {https://doi.org/10.1126/science.1261375} {\bibfield  {journal} {\bibinfo  {journal} {Science}\ }\textbf {\bibinfo {volume} {347}},\ \bibinfo {pages} {262} (\bibinfo {year} {2015})}\BibitemShut {NoStop}%
\bibitem [{\citenamefont {Dweck}(2006)}]{Dweck2006}%
  \BibitemOpen
  \bibfield  {author} {\bibinfo {author} {\bibfnamefont {C.~S.}\ \bibnamefont {Dweck}},\ }\href {https://psycnet.apa.org/record/2006-08575-000} {\emph {\bibinfo {title} {{Mindset: The new psychology of success.}}}}\ (\bibinfo  {publisher} {Random House},\ \bibinfo {year} {2006})\BibitemShut {NoStop}%
\bibitem [{\citenamefont {Kalender}\ \emph {et~al.}(2022)\citenamefont {Kalender}, \citenamefont {Marshman}, \citenamefont {Schunn}, \citenamefont {Nokes-Malach},\ and\ \citenamefont {Singh}}]{Kalendar2022}%
  \BibitemOpen
  \bibfield  {author} {\bibinfo {author} {\bibfnamefont {Z.~Y.}\ \bibnamefont {Kalender}}, \bibinfo {author} {\bibfnamefont {E.}~\bibnamefont {Marshman}}, \bibinfo {author} {\bibfnamefont {C.~D.}\ \bibnamefont {Schunn}}, \bibinfo {author} {\bibfnamefont {T.~J.}\ \bibnamefont {Nokes-Malach}},\ and\ \bibinfo {author} {\bibfnamefont {C.}~\bibnamefont {Singh}},\ }\bibfield  {title} {\bibinfo {title} {{Framework for unpacking students' mindsets in physics by gender}},\ }\href {https://doi.org/10.1103/PhysRevPhysEducRes.18.010116} {\bibfield  {journal} {\bibinfo  {journal} {Physical Review Physics Education Research}\ }\textbf {\bibinfo {volume} {18}},\ \bibinfo {pages} {010116} (\bibinfo {year} {2022})}\BibitemShut {NoStop}%
\bibitem [{\citenamefont {Scherr}\ \emph {et~al.}(2017)\citenamefont {Scherr}, \citenamefont {Plisch}, \citenamefont {Gray}, \citenamefont {Potvin},\ and\ \citenamefont {Hodapp}}]{Scherr2017}%
  \BibitemOpen
  \bibfield  {author} {\bibinfo {author} {\bibfnamefont {R.~E.}\ \bibnamefont {Scherr}}, \bibinfo {author} {\bibfnamefont {M.}~\bibnamefont {Plisch}}, \bibinfo {author} {\bibfnamefont {K.~E.}\ \bibnamefont {Gray}}, \bibinfo {author} {\bibfnamefont {G.}~\bibnamefont {Potvin}},\ and\ \bibinfo {author} {\bibfnamefont {T.}~\bibnamefont {Hodapp}},\ }\bibfield  {title} {\bibinfo {title} {{Fixed and growth mindsets in physics graduate admissions}},\ }\href {https://doi.org/10.1103/PhysRevPhysEducRes.13.020133} {\bibfield  {journal} {\bibinfo  {journal} {Physical Review Physics Education Research}\ }\textbf {\bibinfo {volume} {13}},\ \bibinfo {pages} {020133} (\bibinfo {year} {2017})}\BibitemShut {NoStop}%
\bibitem [{\citenamefont {Little}\ \emph {et~al.}(2019)\citenamefont {Little}, \citenamefont {Humphrey}, \citenamefont {Green}, \citenamefont {Nair},\ and\ \citenamefont {Sawtelle}}]{Little2019}%
  \BibitemOpen
  \bibfield  {author} {\bibinfo {author} {\bibfnamefont {A.~J.}\ \bibnamefont {Little}}, \bibinfo {author} {\bibfnamefont {B.}~\bibnamefont {Humphrey}}, \bibinfo {author} {\bibfnamefont {A.}~\bibnamefont {Green}}, \bibinfo {author} {\bibfnamefont {A.}~\bibnamefont {Nair}},\ and\ \bibinfo {author} {\bibfnamefont {V.}~\bibnamefont {Sawtelle}},\ }\bibfield  {title} {\bibinfo {title} {{Exploring mindset's applicability to students' experiences with challenge in transformed college physics courses}},\ }\href {https://doi.org/10.1103/PhysRevPhysEducRes.15.010127} {\bibfield  {journal} {\bibinfo  {journal} {Physical Review Physics Education Research}\ }\textbf {\bibinfo {volume} {15}},\ \bibinfo {pages} {010127} (\bibinfo {year} {2019})}\BibitemShut {NoStop}%
\end{thebibliography}

%

\end{document}